\documentclass[aps,prc,twocolumn,superscriptaddress]{revtex4-1}

\usepackage[english]{babel}
\usepackage{graphicx}
\usepackage{color}
\usepackage{amssymb}
\usepackage{amsmath}
\usepackage{enumitem}
\usepackage{array}
\usepackage{booktabs}

\renewcommand{\vec}[1]{\mbox{\boldmath$#1$}}

\newcommand{\Tmn}{T^{\mu \nu}}
\newcommand{\jmu}{j^{\mu}}

\begin{document}

\title{Deviations of the Energy-Momentum Tensor from Equilibrium in the Initial State for Hydrodynamics from Transport Approaches}

\author{D.~Oliinychenko}
\email{oliiny@fias.uni-frankfurt.de}
%\thanks{}
\affiliation{Frankfurt Institute for Advanced Studies, D-60438
  Frankfurt am Main, Germany}
\affiliation{Bogolyubov Institute for Theoretical
  Physics, Kiev 03680, Ukraine}

\author{H.~Petersen}
\email{petersen@fias.uni-frankfurt.de}
\affiliation{Frankfurt
  Institute for Advanced Studies, D-60438 Frankfurt am Main, Germany}
\affiliation{Institut f\"ur Theoretische Physik,
  Goethe-Universit\"at, D-60438 Frankfurt am Main, Germany}
\affiliation{GSI Helmholtzzentrum f\"ur Schwerionenforschung GmbH,
  Planckstr. 1, 64291 Darmstadt, Germany}

\begin{abstract}
Many hybrid models of heavy ion collisions construct the initial state for hydrodynamics from transport models. Hydrodynamics requires that the energy-momentum tensor $\Tmn$ and four-currents $\jmu$ do not deviate considerably from the equilibrium ideal-fluid form, but the ones constructed from transport do not necessarily possess this property. In this work we investigate the space-time picture of $\Tmn$ deviations from equilibrium in Au+Au collisions using a coarse-grained transport approach. The collision energy is varied in the range $E_{lab} = 5-160\emph{A}$ GeV. The sensitivity of $\Tmn$ deviations from equilibrium to collision centrality, and other parameters such as the switching criterion, the amount of statistics used to construct the initial state, and the smearing parameter $\sigma$ is investigated. For low statistics deviations of $\Tmn$ from equilibrium are large and dominated by the effect of finite sampling. For large statistics the pressure anisotropy plays the most significant role, while the off-diagonal components of $\Tmn$ are small in a large volume during the whole evolution. For all considered energies and centralities the pressure anisotropy exhibits a similar feature: there is a narrow interval of time, when it rapidly drops in a considerable volume. This allows us to introduce an ''isotropization time'', which is found to decrease with energy and slightly increase with centrality. The isotropization times are larger than times typically used for initializing hydrodynamics.
\end{abstract}

\maketitle

\section{Introduction}

Heavy ion collision experiments at the Relativistic Heavy Ion Collider (RHIC) at center of mass energy $\sqrt{s_{NN}} = 200$~GeV per nucleon pair at Brookhaven and at the Large Hadron Collider (LHC) at Geneva at $\sqrt{s_{NN}} = 2.76$~TeV have provided evidence of the Quark-Gluon Plasma (QGP) existence and proven that it behaves almost like a perfect fluid \cite{Arsene:2004fa,Adcox:2004mh,Back:2004je,Adams:2005dq,Muller:2012zq,Shuryak:2004cy}. Models based on ideal hydrodynamics, which requires local thermalization, are very successful in describing the experimental data from RHIC and LHC. It was argued that the good description of experimental data requires rather small initialization times ($\tau \leq 0.6$ fm/c at RHIC) \cite{Heinz:2004pj}. This raised the question about how fast and due to which physical mechanisms the QGP can be thermalized. In contrast to this requirement of early thermalization, in the newer works applying viscous hydrodynamics it is claimed that ''experimental data for bulk quantities can be reproduced by hydrodynamic models also for large initialization times, so no early thermalization assumption is needed'' \cite{Luzum:2008cw}. Still, the necessary condition for applying hydrodynamics is the vicinity to the thermal equilibrium. The recently developed anisotropic hydrodynamics \cite{Strickland:2014pga} is partially an exception: it can be applied for arbitrarily high pressure anisotropy, but not for arbitrarily high off-diagonal components of the energy-momentum tensor. Unlike ideal or viscous hydrodynamics, anisotropic hydrodynamics is not yet a well-established ansatz, for a discussion see \cite{Naboka:2014eha}.

The question about the degree of non-equilibrium at the time of hydrodynamics initialization cannot be answered by hydrodynamics itself. The approach to equilibrium in heavy ion collisions is actively studied microscopically within two types of models, an overview of both being given in \cite{Strickland:2013uga}. The first ones apply dualities of supersymmetric Yang-Mills gauge theory for calculations in the strong coupling limit \cite{Heller:2011ju,vanderSchee:2013pia}. Other approaches are able to achieve fast thermalization in a weak coupling limit, where colliding nuclei are described in the color-glass condensate (CGC) framework \cite{Gelis:2010nm,Lappi:2010ek}. The primary effect in CGC leading to fast thermalization is believed to be plasma instabilities, such as the chromo-Weibel instability\cite{Arnold:2004ti}. Both kinds of models predict considerable momentum space anisotropies at the times, when hydrodynamics is initialized. The aforementioned studies are relevant for high collision energies. At intermediate energies thermalization was studied using transport models \cite{Bravina:2008ra}, where momentum distributions were averaged over a ($5 \times 5 \times 5$) fm$^3$ central cell. In contrast to the work \cite{Bravina:2008ra}, in this paper we study thermalization and isotropization at intermediate energies locally in space, paying attention not only to pressure anisotropy, but also to off-diagonal components of the energy-momentum tensor. Our study is particularly relevant for hydrodynamics+transport hybrid models that construct initial state from transport approaches \cite{Petersen:2008dd,Skokov:2005ut,Werner:2010aa,Andrade:2005tx,Gale:2012rq,Karpenko:2015xea,Pang:2012he}.

In such models the initial energy-momentum tensor $\Tmn$ and the four-current $\jmu$ are obtained from particles, a procedure we further refer to as \emph{fluidization}. Afterwards they can be unambiguously decomposed into $\Tmn = \Tmn_{ideal} + \tau^{\mu \nu}$ and $\jmu = \jmu_{ideal} + q^{\mu}$, where $\Tmn_{ideal}$ and $\jmu_{ideal}$ are the energy-momentum tensor and the four-current of conserved charges of an ideal fluid. For a fully consistent fluidization a set of conditions is required. First, the hydrodynamics should be applicable, therefore the system has to be close to equilibrium.
\begin{align} 
||\tau^{\mu \nu}|| \ll ||\Tmn_{ideal}|| \label{EqI} \\
||q^{\mu}|| \ll ||\jmu_{ideal}|| \label{EqII}
\end{align}
For applicability of hydrodynamics it is also required that the Knudsen number is small,
\begin{align}
Kn \equiv l_{micro}/L_{macro} \ll 1 \,, \label{EqIII}
\end{align} 
where $l_{micro}$ is a microscopic scale related to particle collisions, e.g. mean free path, and $L_{macro}$ is a macroscopic scale, e.g. the system size. Second, hydrodynamics and transport have to be matched: energy-momentum tensor, currents and their time derivatives should be the same for transport and hydrodynamics at fluidization. Matching time derivatives means that the transport properties (viscosities, thermal conductivity, relaxation time, etc), chemical rates and the equation of state (EoS) should be the same on both sides:
\begin{align}
\Tmn_{transport} = \Tmn_{hydro}\, \label{EqIV} \\
\jmu_{transport} = \jmu_{hydro}\, \label{EqV} \\
\text{same transport properties, chemical rates and EoS}\, \label{EqVI}
\end{align}

Let us call this set of conditions (\ref{EqI}-\ref{EqVI}) \emph{strong consistency} conditions. These conditions guarantee that for a small time after fluidization results of the hydrodynamic calculation are equivalent to the results of the transport model. Hydrodynamics in this case is just another language for the description of the same phenomena, which may have some technical advantages. This is the case for the applications of non-relativistic hydrodynamics (see, e.g., \cite{Tiwari:1998aa} and references therein): for boundary layers and dilute regions, where hydrodynamics is not applicable, a transport model is used, otherwise hydrodynamics is applied, because it is computationally cheaper. In such models simulating the whole system with transport and comparing to the hybrid model is just a consistency check - both approaches should give identical results. Strong consistency conditions are a necessary and sufficient condition for that.

The purpose of applying hydrodynamics in the heavy ion collisions simulations is completely different from the non-relativistic applications, where the hydrodynamics is just a matter of the simulation speed-up. In heavy ion collisions switching to hydrodynamics allows to investigate the effects of thermalization, equation of state and transport properties of the hot and dense hadronic or QGP matter. This implies instantaneous thermalization, rapid change of EoS and transport properties at fluidization. Therefore, conditions (\ref{EqIII}) and (\ref{EqVI}) are not required, but conservation laws represented by equations (\ref{EqIV}-\ref{EqV}) should be fulfilled and conditions (\ref{EqI}-\ref{EqII}) remain as a practical requirement to solve the hydrodynamic equations numerically. We further call conditions (\ref{EqI}-\ref{EqII},\ref{EqIV}-\ref{EqV}) the \emph{weak consistency} conditions.

The fluidization is typically performed either at a constant proper time hypersurface $\tau = const$ or at a constant center of mass frame time hypersurface $t_{CM} = const$. The constant is often chosen according to the \emph{geometrical criterion} - the time, when nuclei geometrically pass through each other: $t_{CM} = \frac{2R}{\gamma \beta} = 2R (E_{lab}/2m_N)^{-1/2}$, where $R$ is radius of the nucleus, $\vec{\beta}$ is velocity, $\gamma = (1 - \vec{\beta}^2)^{-1/2}$, $E_{lab}$ is laboratory frame kinetic energy per nucleon, and $m_N$ is nucleon mass. This time is taken to be the same for all collision centralities. It was never systematically verified, if consistency (strong or weak) is fulfilled at fluidization.

In this contribution we want to close this gap and answer the following questions:
\begin{itemize}[noitemsep]
    \item Which conditions of weak consistency are fulfilled at the time of geometrical overlap?
    \item What is a better fluidization criterion? Should it depend on centrality?
    \item How can one perform fluidization consistently in the weak sense?
\end{itemize}

To answer these questions we analyse the outcome of a hadron cascade and determine, where and when conditions (\ref{EqI}-\ref{EqII}) are fulfilled.

The article is organized as follows. To set the stage, the way fluidization is performed in different models is discussed in Section \ref{Sec_II}. In Section \ref{Sec_III} we derive the expressions to quantify deviations of $\Tmn$ and $\jmu$ from the ideal fluid. The methodology of our work is described in Section \ref{sec:Methodology}. Results for quantifying the deviations are split into two sections: in Section \ref{Sec_V} we consider effects of nuisance parameters: statistics, smearing and grid spacing, while in Section \ref{sec:Results} we highlight the dependence on collision energy and centrality. Our criticism to the smearing procedure in the fluidization of the existing models and a suggestion for its improvement are given in the Appendix.

\section{Overview of fluidization in current hybrid models} \label{Sec_II}

Let us start by summarizing the current approaches that are employed in hybrid models (see Table \ref{Tab:models}). All the shown approaches need to obtain the ideal fluid part of $\Tmn$ and $\jmu$ from discrete degrees of freedom (hadrons, partons, strings). The viscous corrections are neglected in all models, even if viscous hydrodynamics is applied for the evolution. The only exception is a recent work by Liu et al. \cite{Liu:2015nwa}, where the initial stage for viscous hydrodynamics is constructed from free streaming partons and viscous corrections are explicitly included.

The energy-momentum tensor $\Tmn$ and four-currents $\jmu$ are constructed as
\begin{align}
\Tmn_{init}(r) &=& \sum_i \frac{p^{\mu}_i p^{\nu}_i}{p^0_i} K(\vec{r}-\vec{r_i}, \vec{p}) \\
\jmu_{init}(r) &=& \sum_i \frac{p^{\mu}_i}{p^0_i} K(\vec{r}-\vec{r_i}, \vec{p}) \nonumber
\end{align}

Here $K$ is a smearing kernel, which is often taken as a Gaussian in various coordinates. In the Appendix we argue, that $K(\vec{r})d^3r$ should be Lorentz scalar, show that none of the models fully meets this requirement, and derive a simple smearing kernel, which can be employed instead. Note that changing to this Lorentz-invariant kernel leaves the final results rather unaffected.

There are two ways in the literature to match the obtained $\Tmn$ and $\jmu$ to ideal hydrodynamics. The first one is to use only $T^{\mu0}$, $j^0$, assuming that they have ideal fluid form $\Tmn_{ideal} = (\epsilon + p)u^{\mu}u^{\nu} - p g^{\mu \nu}$, $\jmu_{ideal} = n u^{\mu}$, and adding the equation of state (EoS) $p = p(\epsilon,n)$. The following system of equations is then solved (usually iteratively, for details see \cite{Pang:2012he}):
\begin{align}
  \begin{cases}
    T^{00} = (\epsilon + p) \gamma^2 - p \\
    T^{0i} = (\epsilon + p) \gamma^2 \vec{v} \\
    j_B^0 = n \gamma \\
    p = p_{EoS}(n, \epsilon)
  \end{cases}
\end{align}

The advantage of this method is that it conserves energy and momentum. However, this method supports switching only to ideal fluid $\Tmn_{ideal}$, keeping viscous corrections is hardly possible. Even though the switching method conserves energy and momentum, one of the models, which employs it, violates conservation laws \cite{Pang:2012he}, because in \cite{Pang:2012he} the whole $\Tmn$ is multiplied by a free parameter $K$, which is then fixed by experimental multiplicities.  

Another way is to determine the energy density $\epsilon$ and the collective velocity $u^{\mu}$ by solving the eigenvalue problem
\begin{align}
  \Tmn u_{\nu} = \epsilon u^{\mu} \,,
\end{align}
using the fact that $u^{\mu}$ is a timelike eigenvector of $\Tmn$ and satisfies $u^{\mu} u_{\mu} = 1$. Then the density $n$ is computed as $n = \jmu u_{\mu}$. Only after that the pressure is determined from EoS. Note that this way is not equivalent to the previous one: here the collective velocity does not depend on the equation of state. This method conserves energy and momentum only if the viscous corrections are kept. If they are neglected (as in \cite{Werner:2010aa,Andrade:2005tx,Gale:2012rq}) then conservation laws are violated. For a simple example assume that $u^{\mu} = \gamma(1,0,0,v)$. In this case, the energy density in the computational frame is $\epsilon_{comp} = \gamma^2(\epsilon + v^2 T^{33}_{L})$, where $T^{33}_L$ can be split into the ideal fluid pressure and a viscous correction. If the correction is neglected, energy conservation is violated.

The switching criteria are rather similar in most of the approaches: either constant proper time $\tau$ or constant time in the center of mass frame $t_{CM}$, the value being determined by the geometrical criterion. It provides the earliest time, when equilibration could be in principle possible. In \cite{Skokov:2006us} the fluidization time $t_{fl}$ has a better physical motivation: $t_{fl}$ is chosen such that the entropy per baryon does not change any more at $t > t_{fl}$.

\begin{table*}
  \begin{tabular}{p{2.5cm}p{2.5cm}p{2.5cm}p{3cm}p{2.5cm}p{2.5cm}p{2.5cm}}
  \toprule[1.5pt]
   Model      &   Initial condition  & Hydro  & Switching \newline criterion & 
   Smearing \newline kernel & Getting \newline $\Tmn_{ideal}$ \\
   
   \midrule[1pt]
     UrQMD \newline hybrid~\cite{Petersen:2008dd} &
     UrQMD \newline cascade &
     ideal 3+1D,\newline SHASTA &
     $t_{CM} \text{[fm/c]} =$ \newline $max(2R \sqrt{\frac{E_{lab}}{2m_N}}, 1.0)$ &
     Gaussian \newline z-contracted &
     $T^{\mu 0}$, $j^0$ \\
     
     Skokov-Toneev \newline hybrid~\cite{Skokov:2005ut} &
     Quark-Gluon- \newline String-Model &
     ideal 3+1D,\newline SHASTA &
     $t_{CM}$ such \newline that $S/Q_B = \text{const}$ &
     not \newline mentioned &
     $T^{\mu 0}$, $j^0$ \\
   
     EPOS~\cite{Werner:2010aa} &
     Strings (Regge-\newline Gribov model) &
     ideal 3+1D &
     $\tau$ &
     Gaussian \newline z-contracted &
     Landau frame \\
   
     NeXSPheRIO \newline hybrid~\cite{Andrade:2005tx,Drescher:2000ec} &
     Strings (Regge-\newline Gribov model) &
     ideal 3+1D, \newline SPH &
     $\tau = 1$ fm \cite{Hama:2004rr} &
     Gaussian in \newline $x$, $y$, $\tau \eta$&
     Landau frame \\
   
     Gale et al~\cite{Gale:2012rq} &
     IP-glasma &
     viscous 3+1D,\newline MUSIC &
     $\tau = 0.2$ fm/c \newline ($\sqrt{s_{NN}} = 2.76$ TeV) &
     not \newline mentioned &
     Landau frame \\
   
     Karpenko \newline hybrid~\cite{Karpenko:2015xea} &
     UrQMD \newline cascade &
     viscous 3+1D &
     $\tau_{geom}$ &
     Gaussian with \newline $\sigma_{\perp}$ and $\sigma_{\eta}$ &
     $T^{\mu 0}$, $j^0$ \\
     
     Pang et al \newline hybrid~\cite{Pang:2012he} &
     AMPT &
     ideal 3+1D, \newline SHASTA &
     $\tau$ &
     Gaussian with \newline $\sigma_{\perp}$ and $\sigma_{\eta}$ &
     $T^{\mu 0}$, $j^0$ \\
     
  \bottomrule[1.5pt]
  \end{tabular}
  \caption{Fluidization features in different hybrid approaches. Each of these models, including those using viscous hydrodynamics, neglects viscous corrections at fluidization.}
  \label{Tab:models}
\end{table*}

\section{Expressions for verification of weak consistency} \label{Sec_III}

Let us rewrite the conditions of weak consistency with hydrodynamics for $\Tmn$ and $\jmu$ in a way convenient for numerical computation. General expressions for $\Tmn$ and $\jmu$ in viscous hydrodynamics (Landau picture) are the following:
\begin{align}  \label{EQ:visc_hyd_Tmn}
 \Tmn = \epsilon_0 u^{\mu} u^{\nu} - \Delta^{\mu \nu} (P_0 + \Pi) + \pi^{\mu \nu} \\
 \jmu = n_0 u^{\mu} + q^{\mu} \,, \nonumber
\end{align}
where $\Pi$ is the bulk pressure, $\pi^{\mu \nu}$ is the shear stress tensor, $n_0$ is the conserved quantum number density and $q^{\mu}$ is the diffusion current. Viscous corrections to ideal hydrodynamical $\Tmn$ and $\jmu$ are supposed to be small:
\begin{align}
\label{pi_mn_visc_applic}
||\pi^{\mu \nu}|| \ll ||\Tmn|| \\
\Pi \ll P_0 \\
||q^{\mu}|| \ll n_0
\end{align}
From Eqns. \ref{EQ:visc_hyd_Tmn} one obtains
\begin{align}
\pi^{\mu \nu} = T^{\mu \nu} - \epsilon_0 u^{\mu} u^{\nu} + \frac{1}{3} \Delta^{\mu \nu} (T^{\alpha}_{\alpha} -\epsilon_0) \\
P_0 + \Pi = - \frac{1}{3} \Delta_{\mu \nu} \Tmn \\
q^{\mu} = \Delta^{\mu}_{\nu} j^{\nu}
\end{align}
One can see that in the Landau rest frame $u_L^{\mu} = \operatorname{diag}(1, 0,0,0)$, $\pi_L^{\mu 0} = 0$, and $q_L^{0} = 0$. The non-zero components are written as follows:
\begin{align}
P_0 + \Pi = \frac{1}{3} (T_L^{11} + T_L^{22} + T_L^{33}) \\
\pi^{i j}_L = T^{i j}_L - (P_0 + \Pi)\delta^{ij} \\
q_L^{i} = -j_L^i
\end{align}
Let us note that tensor and vector norms are frame-independent, so the consistency conditions for viscous hydrodynamics can be formulated in any frame. In Eqn. (\ref{pi_mn_visc_applic}) one can substitute $||\Tmn||$ by its largest component in the Landau frame: $\epsilon_0$. Then Eqn. (\ref{pi_mn_visc_applic}) will turn into
\begin{align}
||\Tmn_L - \operatorname{diag}(\epsilon_0, P', P', P') || \ll \epsilon_0 \,,
\end{align}
where $P'$ denotes $\frac{1}{3} (T_L^{11} + T_L^{22} + T_L^{33}) = P_0 + \Pi$. The physical meaning of this equation is that the diagonal components of $\Tmn$ in the Landau rest frame do not deviate much from $P'$ and simultaneously off-diagonal components are small compared to $\epsilon_0$. The condition for $q^{\mu}$ is rewritten as
\begin{align}
(j_L^1)^2 + (j_L^2)^2 + (j_L^3)^2 \ll (j_L^0)^2 \,.
\end{align}
Here the physical meaning is that relative velocity between Landau and Eckart frames should be small.
To rewrite $\Pi \ll P_0$ one has to add an equation of state $P_0 = p_{EoS}(\epsilon_0, n_0)$ to the system. Then one obtains $P'/p_{EoS}(\epsilon_0, j^0_L) - 1\ll 1$. Consequently, whether the tensor $\Tmn$ is suitable for fluid dynamics or not is also defined by the equation of state from the fluid dynamics itself. The same $\Tmn$ can be consistent with viscous hydrodynamics with some equation of state, and may fail when the equation of state is changed. Therefore, we will not study the smallness of bulk corrections further, but leave this for a future study.

The conditions for smallness of the shear stress tensor can be split into two: pressure isotropy and smallness of off-diagonal elements. One has to note that the Landau frame is defined only up to an arbitrary rotation. Locally one can always choose such coordinates that $T^{12}_L = T^{23}_L = T^{13}_L = 0$. However, our coordinates are the global coordinates of the computational frame and therefore non-diagonal components of the $\Tmn_L$ are in general non-zero. Therefore,

\begin{align}
|T^{11}_L - P'| + |T^{22}_L - P'| + |T^{33}_L - P'| \ll \epsilon_0 \\
|T^{12}_L| + |T^{23}_L| + |T^{13}_L| \ll \epsilon_0
\end{align}
To make these conditions stricter, every term is substituted by the right hand side of the inequality $|T^{11}_L - P'| = |T^{11}_L - T^{22}_L + T^{11}_L - T^{33}_L|/3 \le |T^{11}_L - T^{22}_L|/3 + |T^{11}_L - T^{33}_L|/3$ and $\epsilon_0$ is substituted by by $P'$. In this way a set of criteria is obtained that we use for numerical calculations.
\begin{align} \label{cond_to_check}
  X \equiv \frac{|T^{11}_L - T^{22}_L| + |T^{22}_L - T^{33}_L| + |T^{33}_L - T^{11}_L|}{T^{11}_L + T^{22}_L + T^{33}_L} \ll 1 \\
  Y \equiv \frac{3(|T^{12}_L| + |T^{23}_L| + |T^{13}_L|)}{T^{11}_L + T^{22}_L + T^{33}_L} \ll 1 \nonumber \\
  v_{LE} = \sqrt{(j_L^1)^2 + (j_L^2)^2 + (j_L^3)^2}/j_L^0 \ll 1 \nonumber \\
  Z \equiv \frac{T^{11}_L + T^{22}_L + T^{33}_L}{3 \, p_{EoS}(\epsilon_0, j^0_L)} - 1 \ll 1 \nonumber
\end{align}

In the following $X$ is referred to as pressure anisotropy and $Y$ as off-diagonality. Please note that due to the inequality $|a-b| \le |a| + |b|$ it is always fulfilled that $X \le 2$. For ideal fluid dynamics $X = 0$. For $Y$ let us remark that
\begin{align}
T^{12} \sim \sum \frac{p^1 p^2}{p^0} \le \frac{1}{2} \sum \frac{p_1^2 + p_2^2}{p^0} \sim \frac{1}{2} (T^{11} + T^{22})
\end{align}
Interchanging indices and substituting this into the definition of $Y$ one gets $Y \le 3$. For an ideal fluid $Y=0$.

\section{Methodology} \label{sec:Methodology}

Our calculation is based on the hadronic transport approach -
Ultrarelativistic Quantum Molecular Dynamics (UrQMD 3.4)
\cite{ref:UrQMD}. The degrees of freedom in UrQMD are hadrons,
resonances up to a mass of 2.2 GeV and strings and the implemented
processes include binary elastic and inelastic scatterings which
mainly proceed via resonance formation and decays or string excitation
and fragmentation at higher collision energies. The UrQMD particles
move along classical trajectories and scatter according to their
free-particle cross-sections. In our studies there are no long range
potentials and particle trajectories between collisions are always
straight lines. Using UrQMD we simulate Au + Au collisions at
laboratory frame energies $E_{\rm lab} =$ 5, 10, 20, 40, 80 and 160$A$
GeV. 

The general procedure for our calculations is:
\begin{enumerate}[noitemsep]
\item Generate many UrQMD events and coarse-grain them using a 2+1D
  space-time grid. The space dimensions are chosen to be the event plane xz. We choose a 2+1D grid and not 3+1D, because
  observing the behaviour of some quantity on 2D surface versus time
  is much easier and informative for a human than observing quantity on a 3D grid.
  Additionally, in central collisions due to symmetry the event plane
  completely characterises all the volume. The center of mass frame is used as the computational frame in the simulation.
  In all the following text the time is measured in the center of mass frame.
  By the UrQMD convention $t = 0$ is the moment when the contracted spheres of the nuclear radius first touch each other in a central collision.
\item Use particles from the generated events to construct the energy-momentum tensor
  $\Tmn(t,x,z)$ locally for each grid cell. To compute $\Tmn$ we take only participants into account,
  i.e. the particles that took part in at least one collision. To construct $\Tmn$ Gaussian smearing is
  employed, for details see Appendix. We argue that the construction of $\Tmn$ in the existing models is not a Lorentz-invariant  
  procedure due to the smearing kernels. A simple kernel is suggested, which satisfies the necessary Lorentz-transformation
  requirements. However, the introduction of the new kernel appeared to be only a matter of physical rigour:
  switching standard kernel to our kernel did not produce noticeable changes in the results. 
\item Transform the constructed energy-momentum tensor 
  in each grid cell to the Landau rest frame.
\item Verify the weak consistency conditions (see Eqn. \ref{cond_to_check}) locally in time and space. We check conditions
  for smallness of pressure anisotropy, off-diagonality, Eckart velocity relative to Landau,
  but the condition for smallness of bulk pressure compared to pressure is left for future studies.
\end{enumerate}

In this procedure we follow exactly the fluidization in hybrid models,
the only addition is that the weak consistency checks are performed. 

\section{Sensitivity to statistics, grid spacing and smearing}\label{Sec_V}

The deviations of $\Tmn$ in the simulation from the ideal fluid $\Tmn_{ideal}$ can have two distinct reasons. The first one is the deviation of the distribution function $\mathit{f}(\vec{r},\vec{p})$ in the transport approach from equilibrium, this reason is referred to as physical. The second reason is statistical: due to finite number of particles in the simulation, the distribution function is not sampled exactly.

\begin{figure}
  \includegraphics[width = 0.23\textwidth]{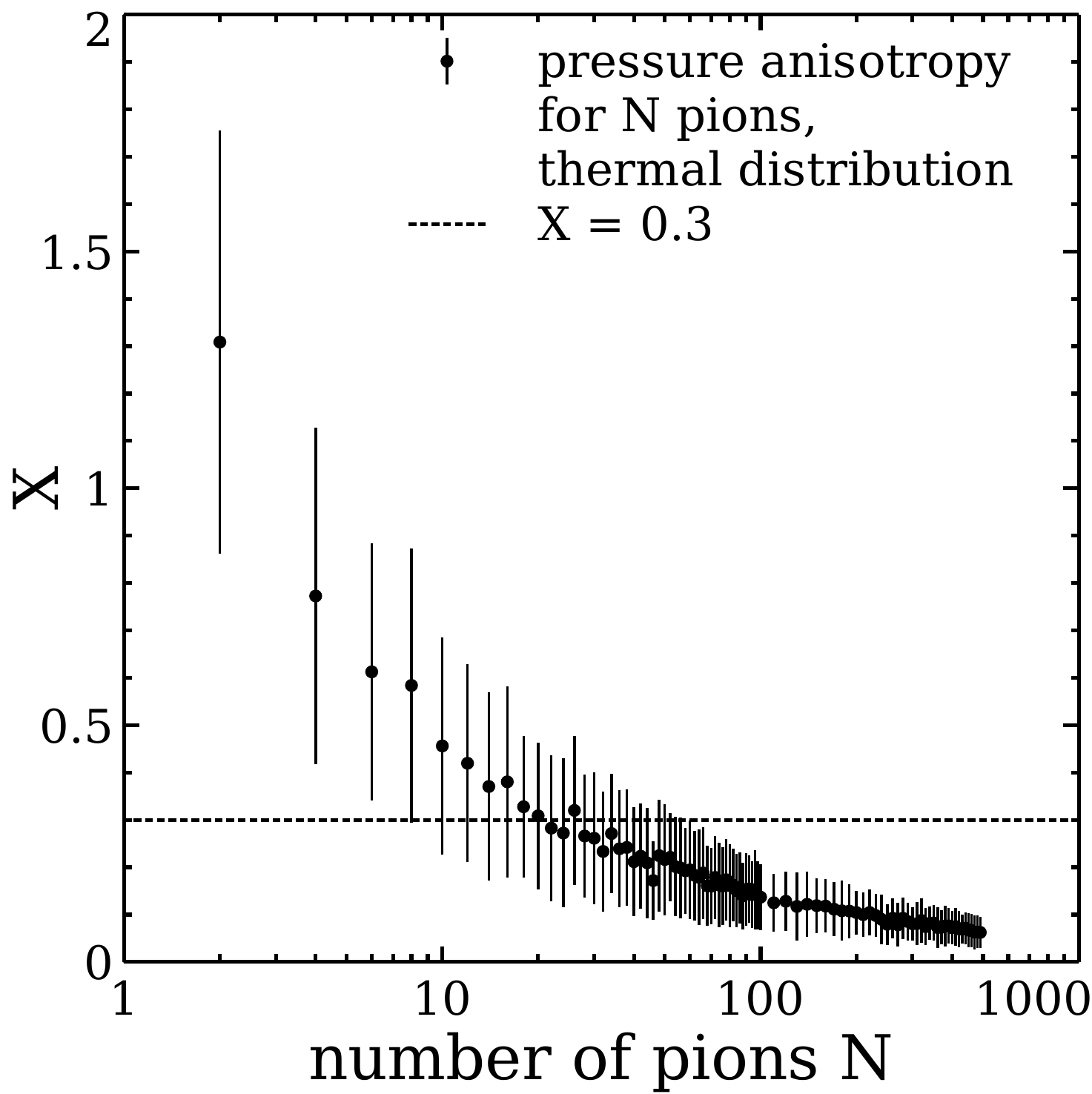}
  \includegraphics[width = 0.23\textwidth]{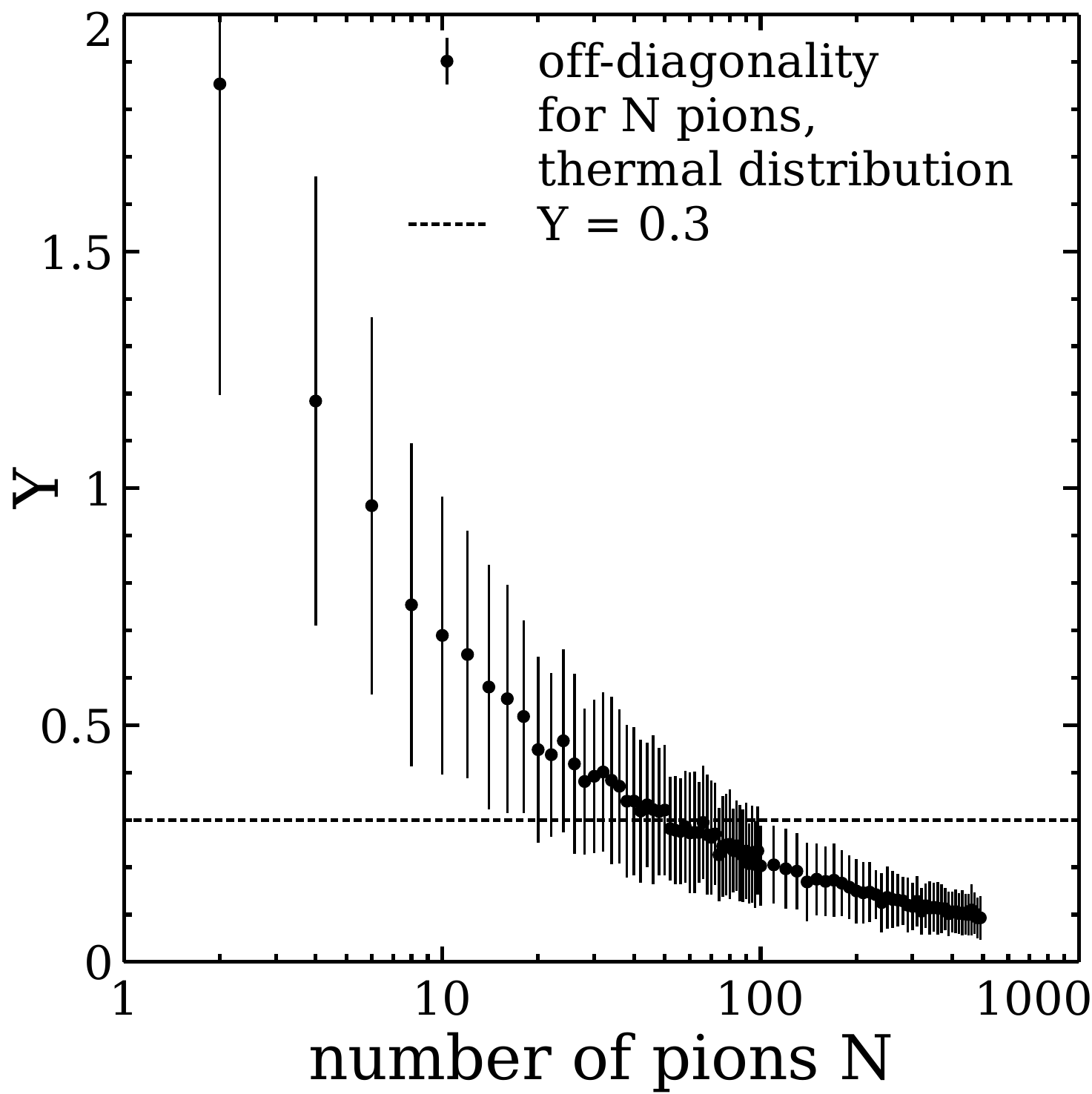}
  \caption{Pressure anisotropy $X$ (left) and off-diagonality $Y$ (right) of $\Tmn$ for particles sampled according to thermal distributions. The effect of statistics on the deviation of energy-momentum tensor from the ideal fluid one is demonstrated.}
  \label{FIG:pure_stat_effect}
\end{figure}

Let us consider two energy-momentum tensors: calculated from particles $\Tmn_{part}$ and a ''true'' $\Tmn$:
\begin{align}
\Tmn_{part}(\vec{r}) =\frac{1}{N_{ev}} \sum_{events} \sum_i \frac{p^{\mu}_i p^{\mu}_i}{p^0_i} K(\vec{r} - \vec{r_i}, p_i) \\
\Tmn(\vec{r}) = \int \frac{p^{\mu} p^{\mu}}{p^0} \mathit{f}(\vec{r},\vec{p}) d^3p
\end{align}
Here $N_{ev}$ is the number of events and $K(\vec{r} - \vec{r_i}, p_i)$ is a smearing kernel. In the limit of $N_{ev} \to \infty$
\begin{align}
\frac{1}{N_{ev}} \sum_{events} \sum_i K(\vec{r} - \vec{r_i}, p_i) \xrightarrow{N_{ev} \to \infty} \\ \int d^3p \, d^3r' \, \mathit{f}(\vec{r'}, \vec{p}) K(\vec{r} - \vec{r'}) \nonumber
\end{align}
For the case of a Gaussian kernel for $\sigma \to 0$, the kernel $K \to \delta(\vec{r} - \vec{r'})$ and one retains the ''true'' $\Tmn$. To combine these two limits ($\sigma \to 0$, $N_{ev} \to \infty$) one has to keep enough particles within a volume of size $\sigma^3$. Consequently, to obtain the ''true'' $\Tmn$ in the simulation, one has to take the limit ($\rho$ is density)
\begin{align} \label{Eq:true_Tmn_condition}
\sigma \to 0,\, N_{ev} \rho \sigma^3 \to \infty
\end{align}
This creates practical limitations for finding the ''true'' $\Tmn$ in simulations: decreasing $\sigma$ 10 times demands increasing the statistics 1000 times! One can also see that regions with lower density are more demanding with respect to statistics. To get some insights into the effect of statistics, we performed an auxiliary simulation: $N$ pions are generated, their momenta being sampled from a thermal distribution with an ad-hoc temperature of $T = 0.2$ GeV, then $\sum \frac{p^{\mu} p^{\mu}}{p^0}$ is computed, pressure isotropy $X$ and off-diagonality $Y$ of the energy-momentum tensor from Eq. \ref{cond_to_check} are calculated. We varied number of pions $N$ and plotted $X$ and $Y$ versus $N$. The results can be seen in Fig. \ref{FIG:pure_stat_effect}. For every point the simulation was repeated 100 times and the standard deviation is displayed as an error.

Fig. \ref{FIG:pure_stat_effect} can be used to specify the number of events needed to reach a good enough approximation to the ''true'' $\Tmn$. For example, for $Y=0.3$ as an acceptable level, the condition of Eqn. \ref{Eq:true_Tmn_condition} becomes $N_{ev} \rho \sigma^3 > 100$. From Fig. \ref{FIG:pure_stat_effect} one can also see that the off-diagonality $Y$ is more sensitive to statistics than the pressure isotropy $X$.

\begin{figure}
  \includegraphics[width = 0.23\textwidth]{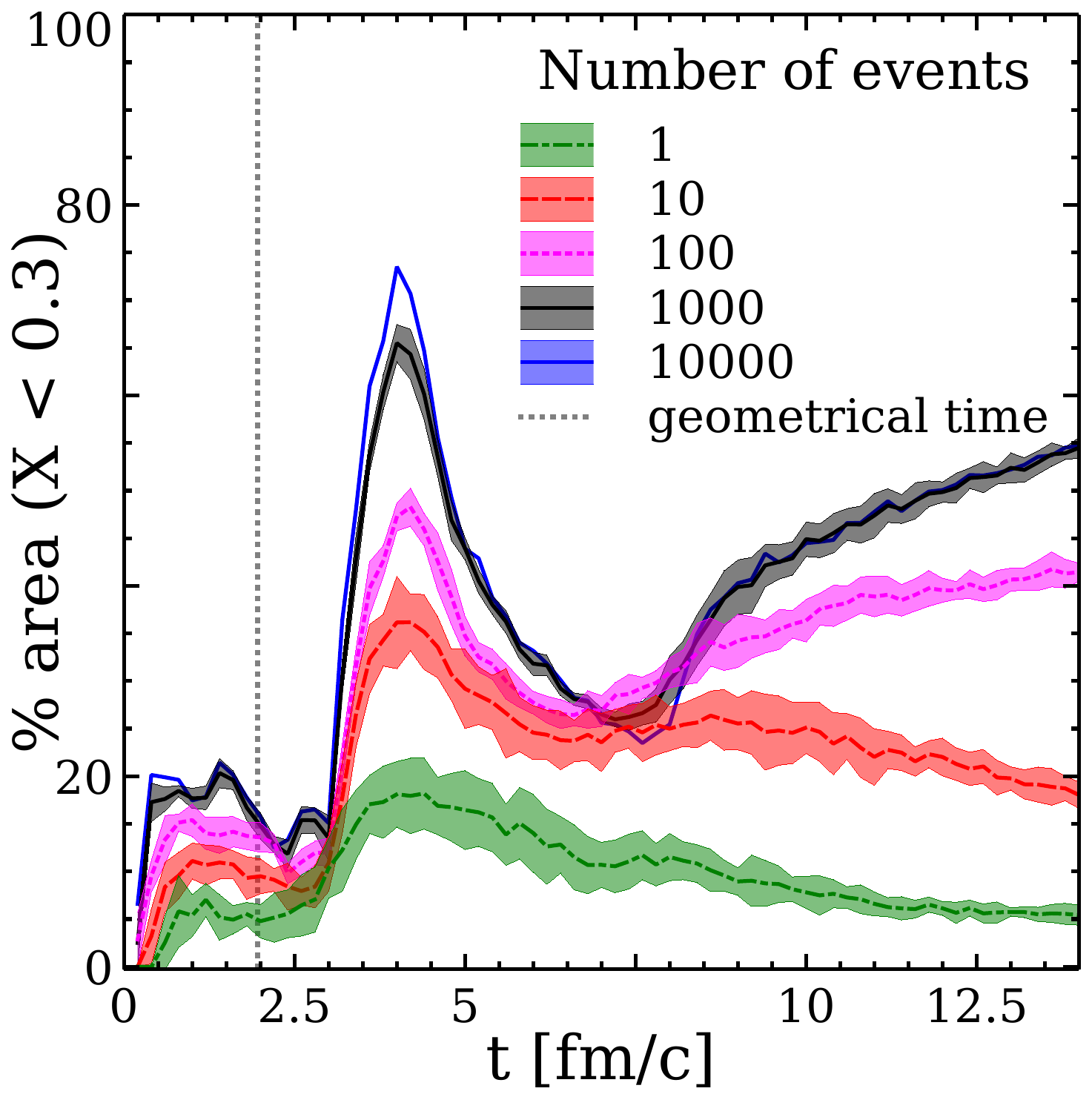}
  \includegraphics[width = 0.23\textwidth]{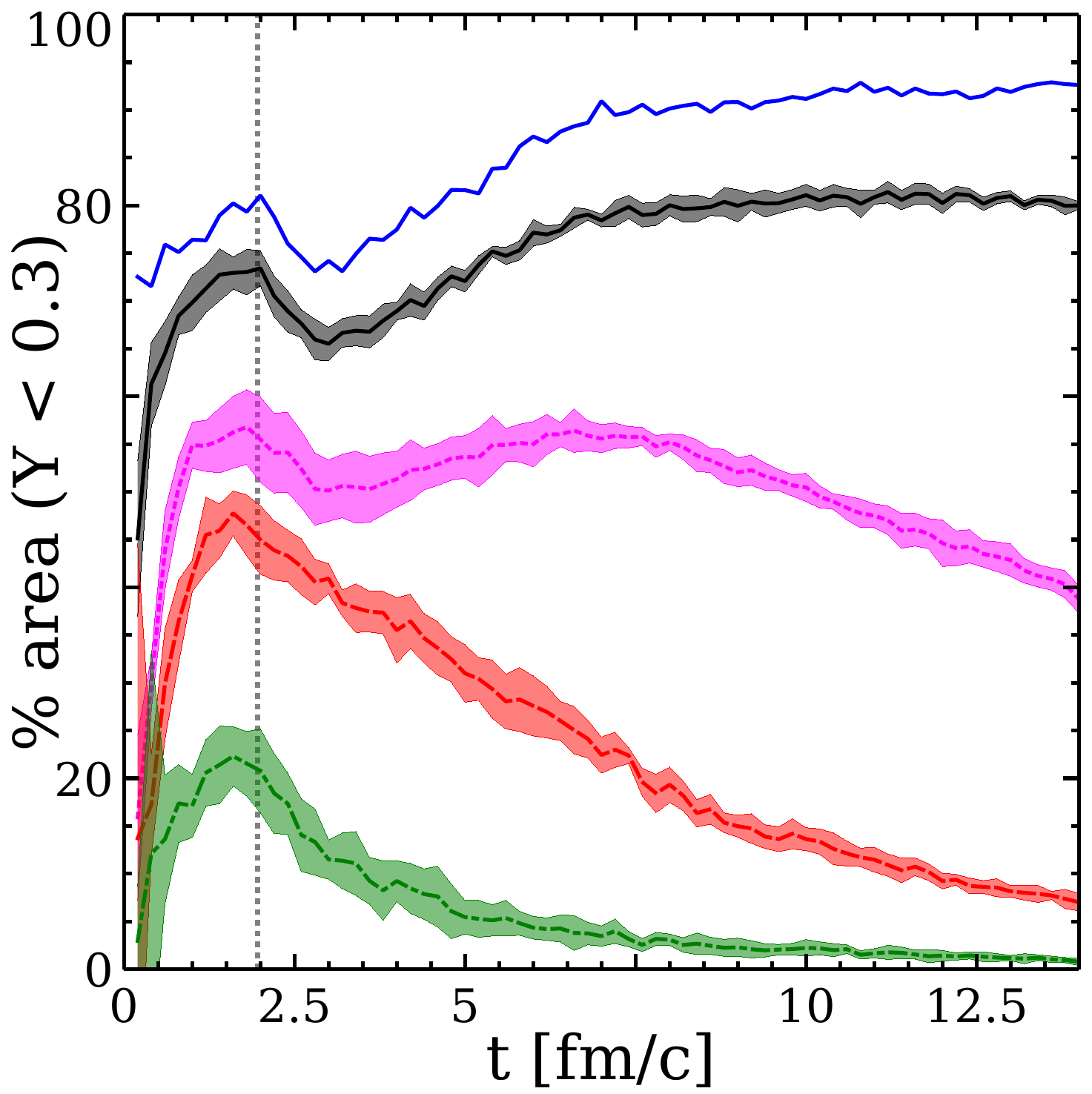}
  \caption{Event plane area percentage, where pressure isotropy $X$ (left) or off-diagonality $Y$ (right) does not exceed 0.3 versus time for different number of events $N_{ev}$ used to construct $\Tmn$. Number of events $N_{ev} = 1$ corresponds to event-by event case.}
  \label{FIG:sensitivity_Nev}
\end{figure}

In the previous paragraph we considered the effect of statistics itself rather as an obstacle to get the physical ''true'' $\Tmn$. However, recently event-by-event simulations gained popularity, where the initial state for hydrodynamics is intentionally constructed from a small number of events to include the fluctuations. Let us see, how the number of events influences deviations of $\Tmn$ from the ideal form in heavy ion collisions. We compute $\Tmn$ locally on every point of the grid, and as a general characteristic we choose the percentage of the event-plane area, where $X < 0.3$ ($Y < 0.3$). To define the total area numerically, only grid cells, where pressure $p > 10^{-4}$ GeV/fm$^3$ are taken into account. For this example Au+Au collisions at $E = 80\emph{A}$ GeV with the impact parameter $b = 6$ fm are considered. The smearing $\sigma$ is 0.8 fm. Results are depicted in Fig.~\ref{FIG:sensitivity_Nev}.

\begin{figure}
  \includegraphics[width = 0.23\textwidth]{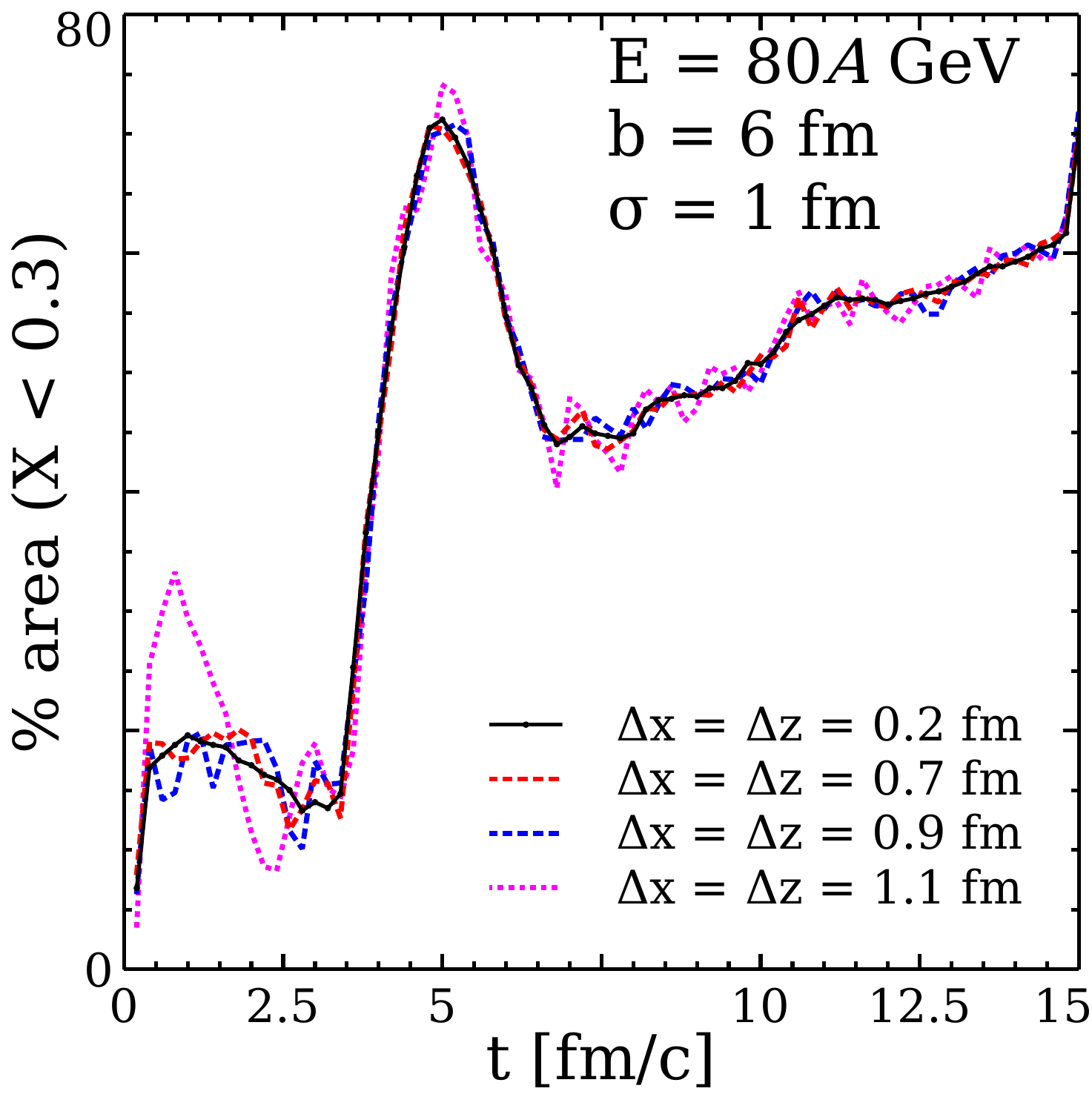}
  \includegraphics[width = 0.23\textwidth]{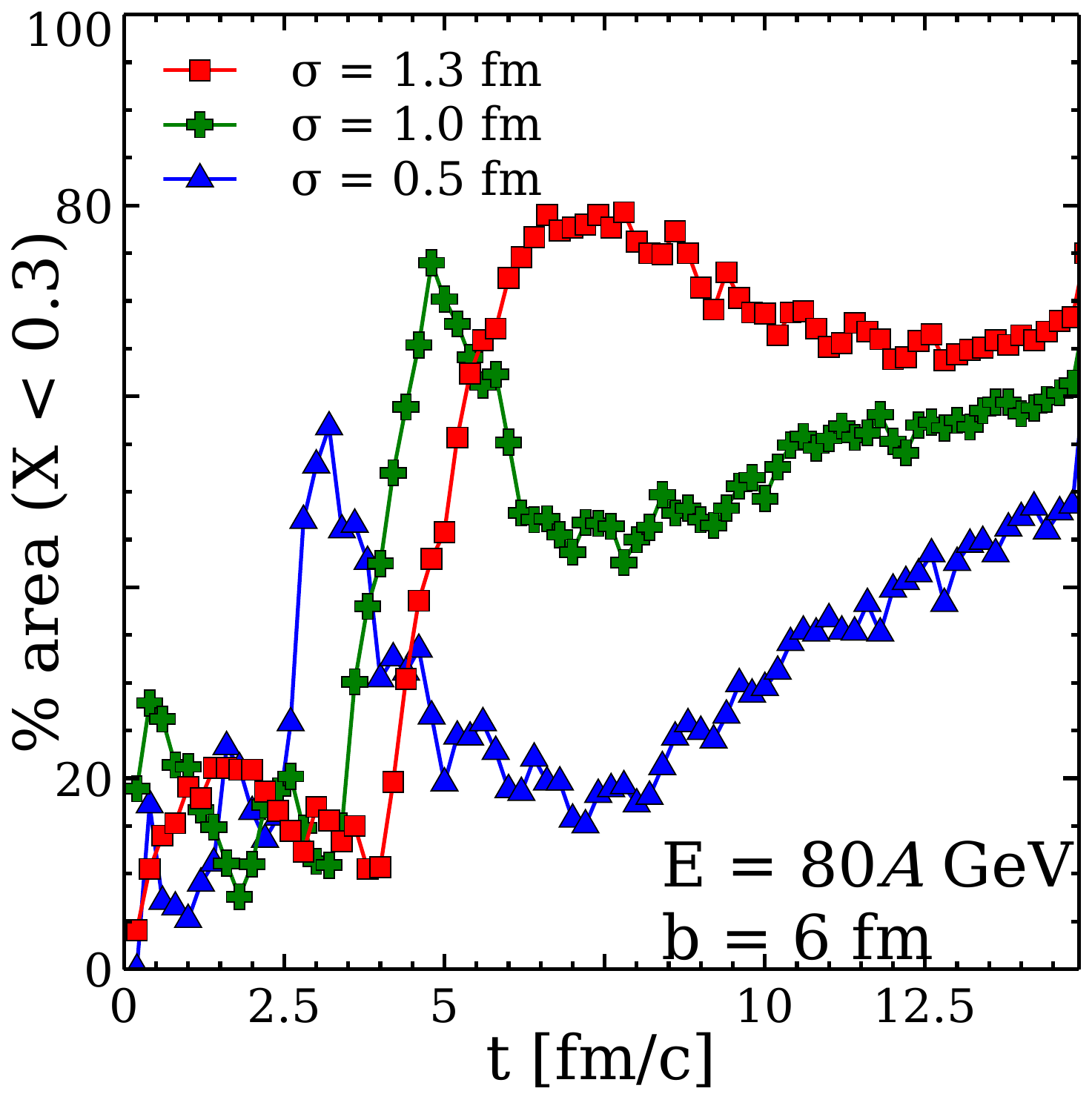}
  \caption{Event plane area percentage, where pressure isotropy $X$ does not exceed 0.3. Au+Au versus time. $E = 80\emph{A}$ GeV, centrality $b = 6$ fm, number of events $N_{ev} = 1000$. Gaussian smearing $\sigma$ (right) and grid spacing $\Delta x = \Delta z$ (left) are varied to study sensitivity of results to them.}
  \label{FIG:sensitivity_sigma_dx}
\end{figure}

\begin{figure}
  \includegraphics[height = 6cm]{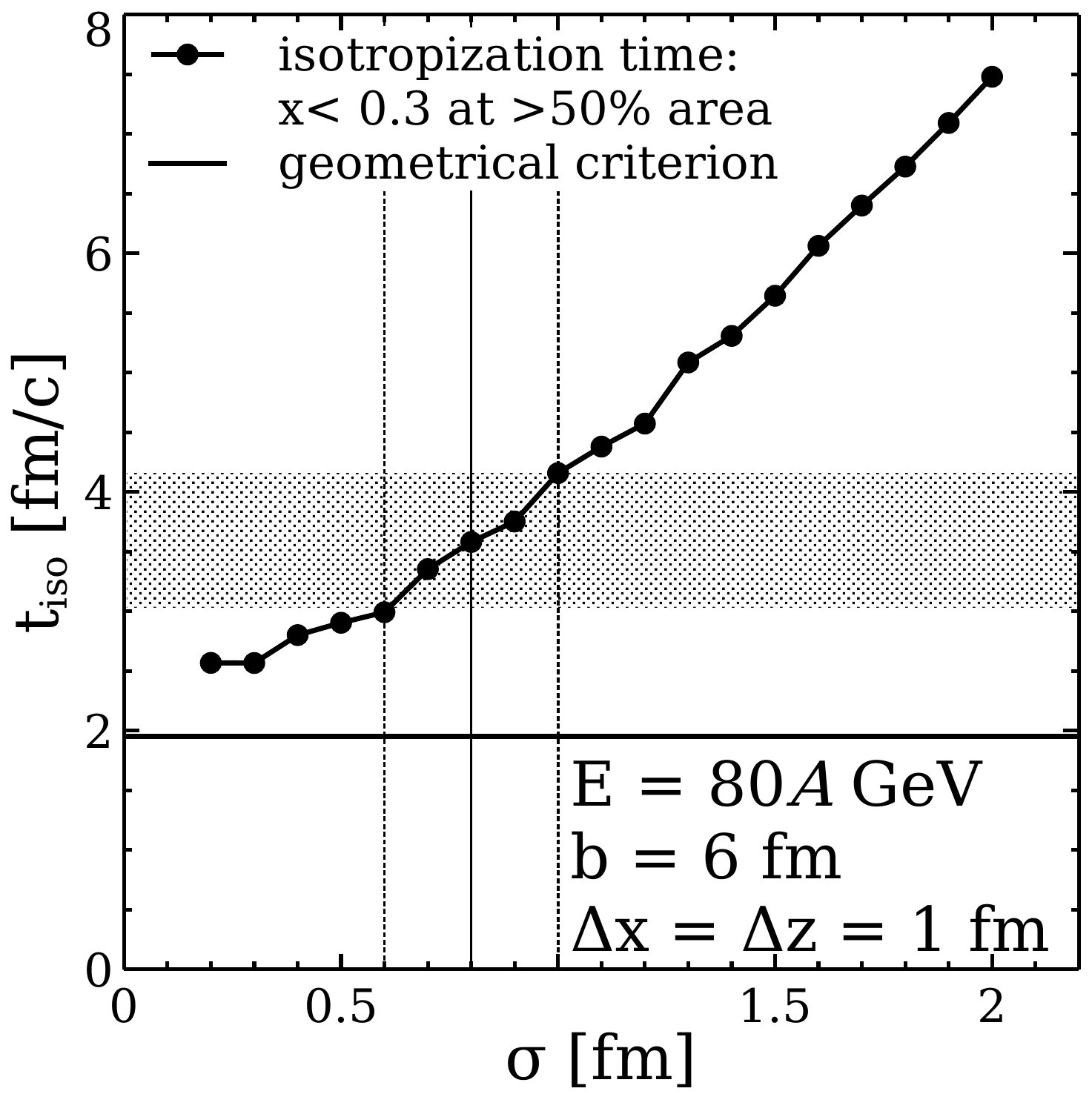}
  \caption{Isotropization time (such time that more than 50\% of event plane area have pressure isotropy $X < 0.3$) versus $\sigma$.}
  \label{FIG:t_iso_sensitivity_sigma}
\end{figure}

One can see that for this given $\sigma$ 1000 events are enough for $X$ to saturate, so the line for $N_{ev} = 1000$ represents results for the physical pressure isotropy, i.e. due to deviation of $\mathit{f}(\vec{r},\vec{p})$ in the transport from equilibrium. For $Y$ at $N_{ev} = 10000$ almost all the event area has small off-diagonality, which means that the physical off-diagonality is small. For event-by-event simulations deviations of $\Tmn$ from ideal fluid are dominated by statistical effects.

In addition to the effect of statistics we investigate the effect of other nuisance parameters, i.e. the grid spacing and the Gaussian smearing, on our results. According to Fig.~\ref{FIG:sensitivity_sigma_dx}, the grid spacing does not influence the results if taken sufficiently small. This is expected, because the grid does not participate in the simulation or in the calculation of $\Tmn$, it only determines the resolution of the $\Tmn$ output. The only effect of grid spacing is on the precision of the area calculation by counting cells, here resolution of the output matters. At early times, it makes some difference, because the total area is small. That is why further we take $\Delta x = \Delta z = 0.6$ fm. At the same time the Gaussian $\sigma$ influences the results very significantly, as it can be seen from Fig.~\ref{FIG:sensitivity_sigma_dx}. The effect of Gaussian $\sigma$ is twofold: on the one hand larger $\sigma$ means effective increase of statistics. On the other hand, if the pressure anisotropy is large at some space point due to physics, the Gaussian smearing will spread this asymmetry in a 1-2 $\sigma$ radius.

\begin{figure*}
  \includegraphics[height = 1cm]{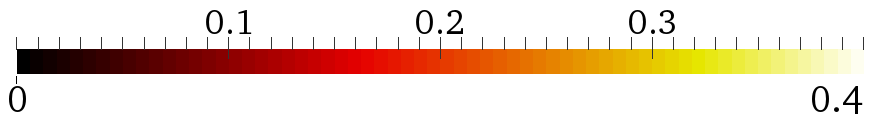} \\
  \includegraphics[width = 3.5cm]{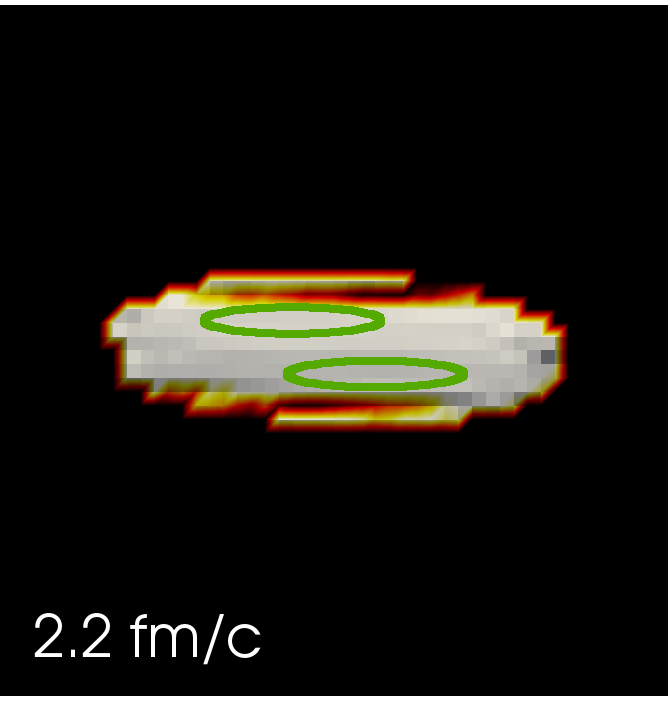}
  \includegraphics[width = 3.5cm]{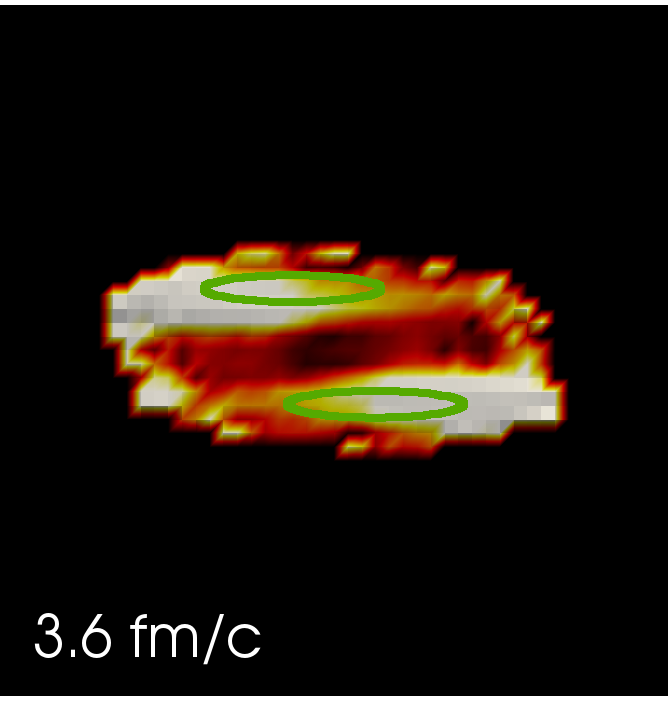}
  \includegraphics[width = 3.5cm]{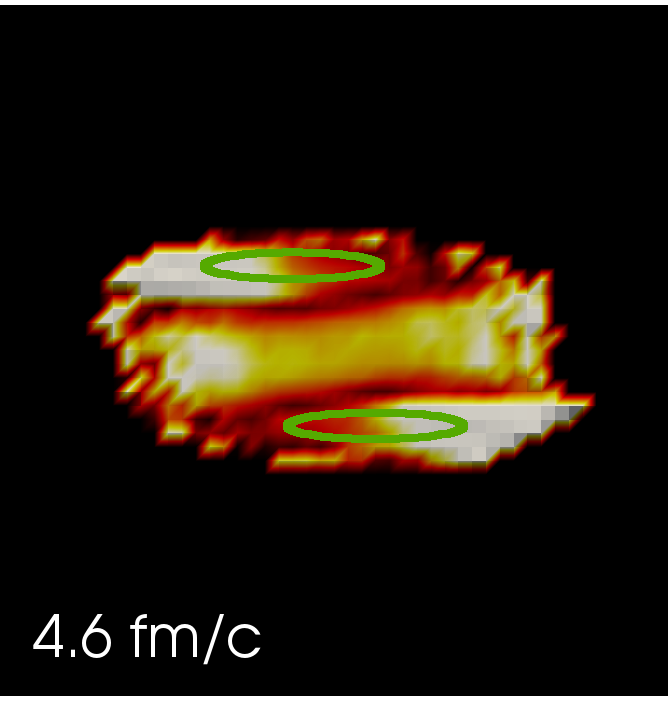}
  \includegraphics[width = 3.5cm]{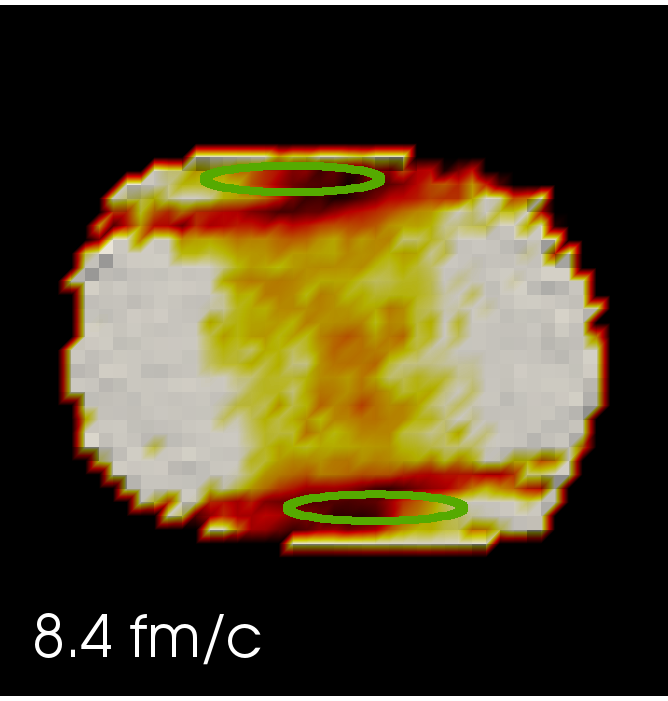}
  \includegraphics[width = 3.5cm]{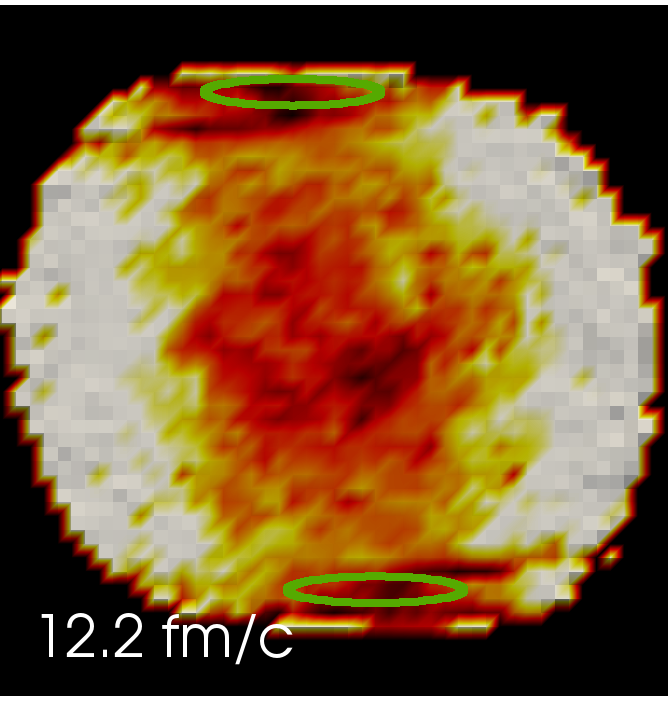}
  \caption{Space-time evolution of pressure anisotropy $X = \frac{|T_L^{11}-T_L^{22}|+|T_L^{22}-T_L^{33}|+|T_L^{33}-T_L^{11}|}{T_L^{11}+T_L^{22}+T_L^{33}}$ (see color scale above the Fig.) for collision energy in lab frame $E = 80\emph{A}$ GeV, centrality $b = 6$ fm. If the value of $X$ exceeds color map maximum, it is marked with the same color as maximum. Solid lines mark the positions of the nuclei, if they wouldn't interact.}
  \label{FIG:x_paraview_space_time_evolution}
\end{figure*}

To characterize the influence of $\sigma$ in a simpler way, we plot $t_{iso}$ versus $\sigma$, where $t_{iso}$ is the earliest time when at least 50\% of the area have $X < 0.3$. We refer to this time as \emph{isotropization time} as it is further described in the following Section. In Fig.~\ref{FIG:t_iso_sensitivity_sigma} this dependence is displayed, the isotropization time is monotonously growing with $\sigma$ and is approaching to the geometrical time for~$\sigma \to 0$. Taking the limit $\sigma \to 0$ is computationally challenging, because one has to increase statistics as $\sigma^{-3}$, as we have shown previously. Instead we choose a reasonable $\sigma = 0.8$ fm and assign systematic errors to our results, corresponding to changing $\sigma$ in the range $(0.6-1.0)$ fm. Another justification for such a treatment is that none of the existing models attempts to consider the ''physical'' limit of $\sigma \to 0,\, N_{ev} \rho \sigma^3 \to \infty$, all the models use some fixed sigma instead.

\section{Results and discussion} \label{sec:Results}

While in the previous section we studied the effects of nuisance parameters on the energy-momentum tensor generated from particles, here we consider the dependence on physical parameters: collision energy and centrality. All the following figures are shown for grid spacing $\Delta x = \Delta z = 0.6$ fm, Gaussian smearing $\sigma = 0.8$ fm, and number of events $N_{ev} = 1000$.

\begin{figure}
  \includegraphics[width = 6cm]{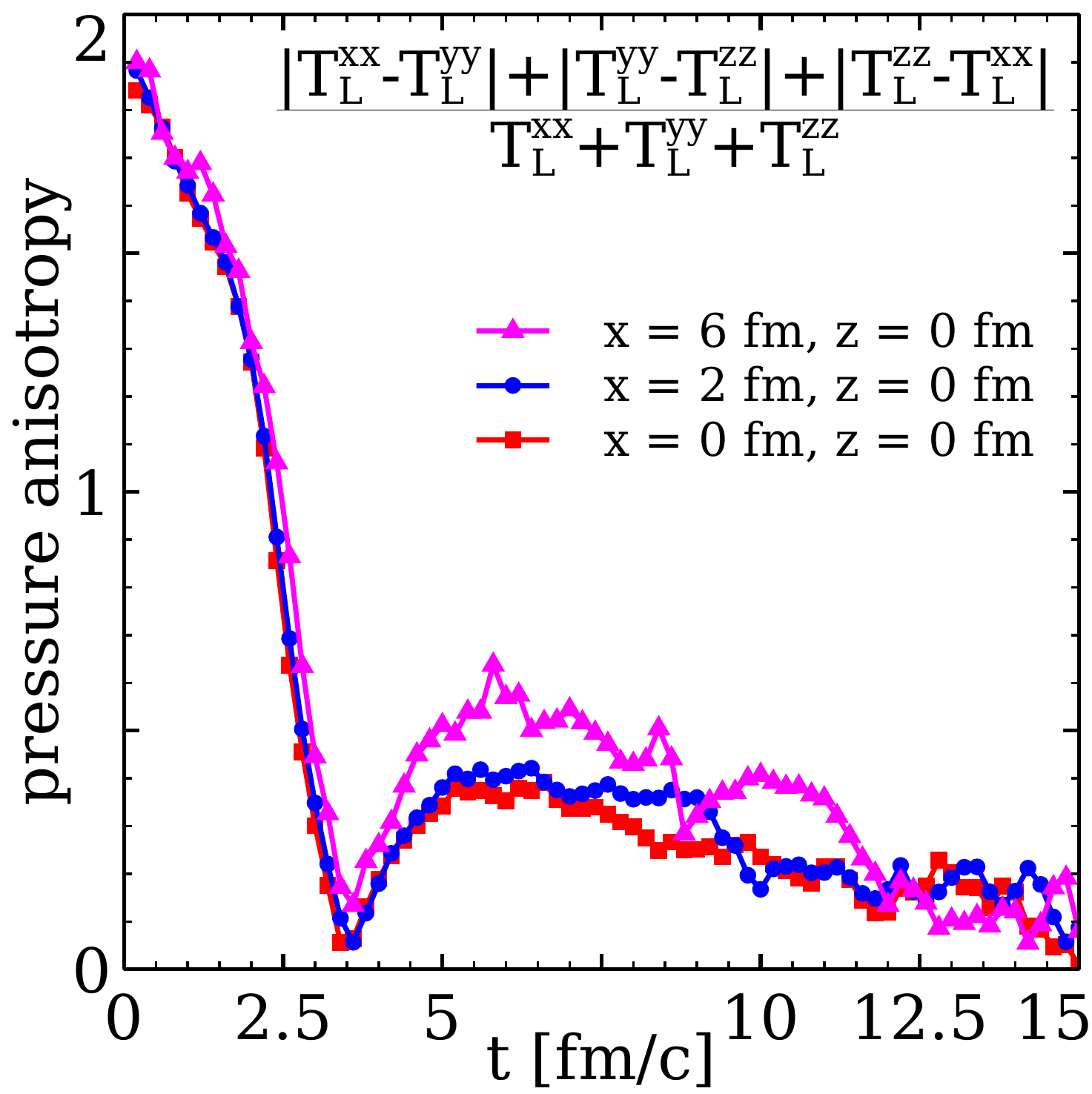}
  \caption{Example of pressure anisotropy behaviour versus time. Collision energy in lab frame
           $E = 80\emph{A}$ GeV, centrality $b = 6$ fm.}
  \label{FIG:x_vs_t_few_space_points}
\end{figure}

\begin{figure}
  \includegraphics[width = 0.2\textwidth]{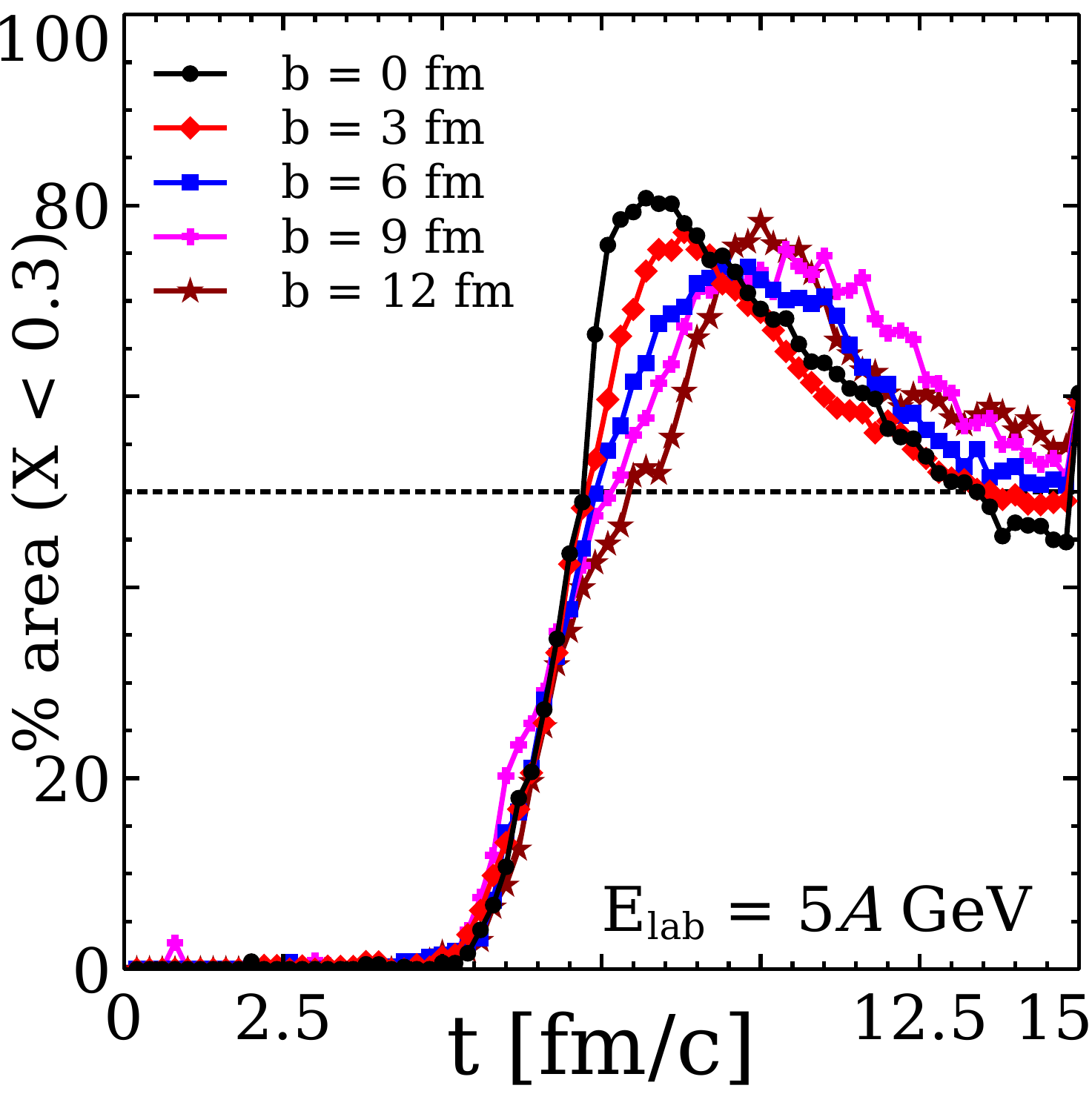}
  \includegraphics[width = 0.2\textwidth]{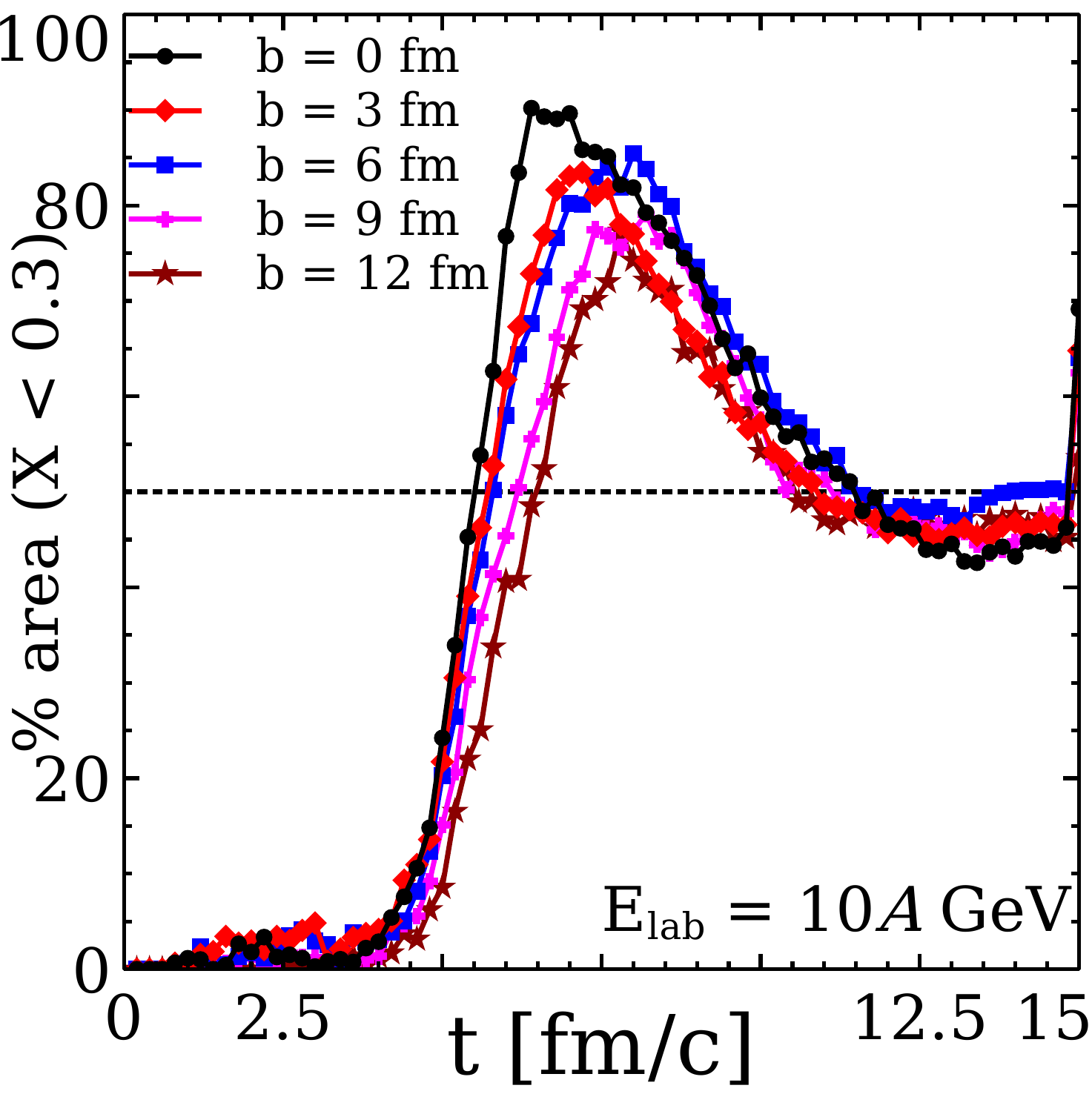}\\
  \includegraphics[width = 0.2\textwidth]{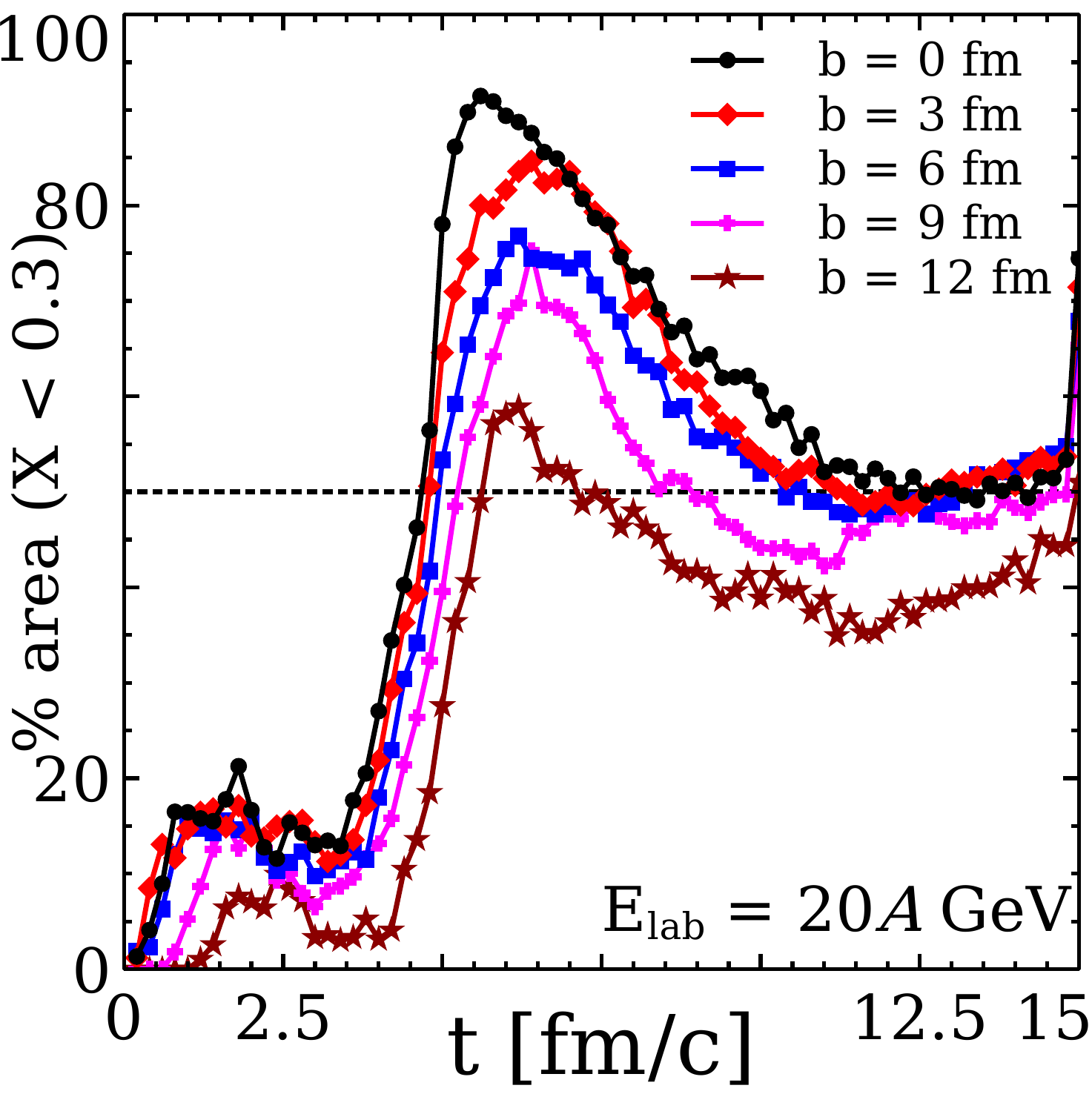}  
  \includegraphics[width = 0.2\textwidth]{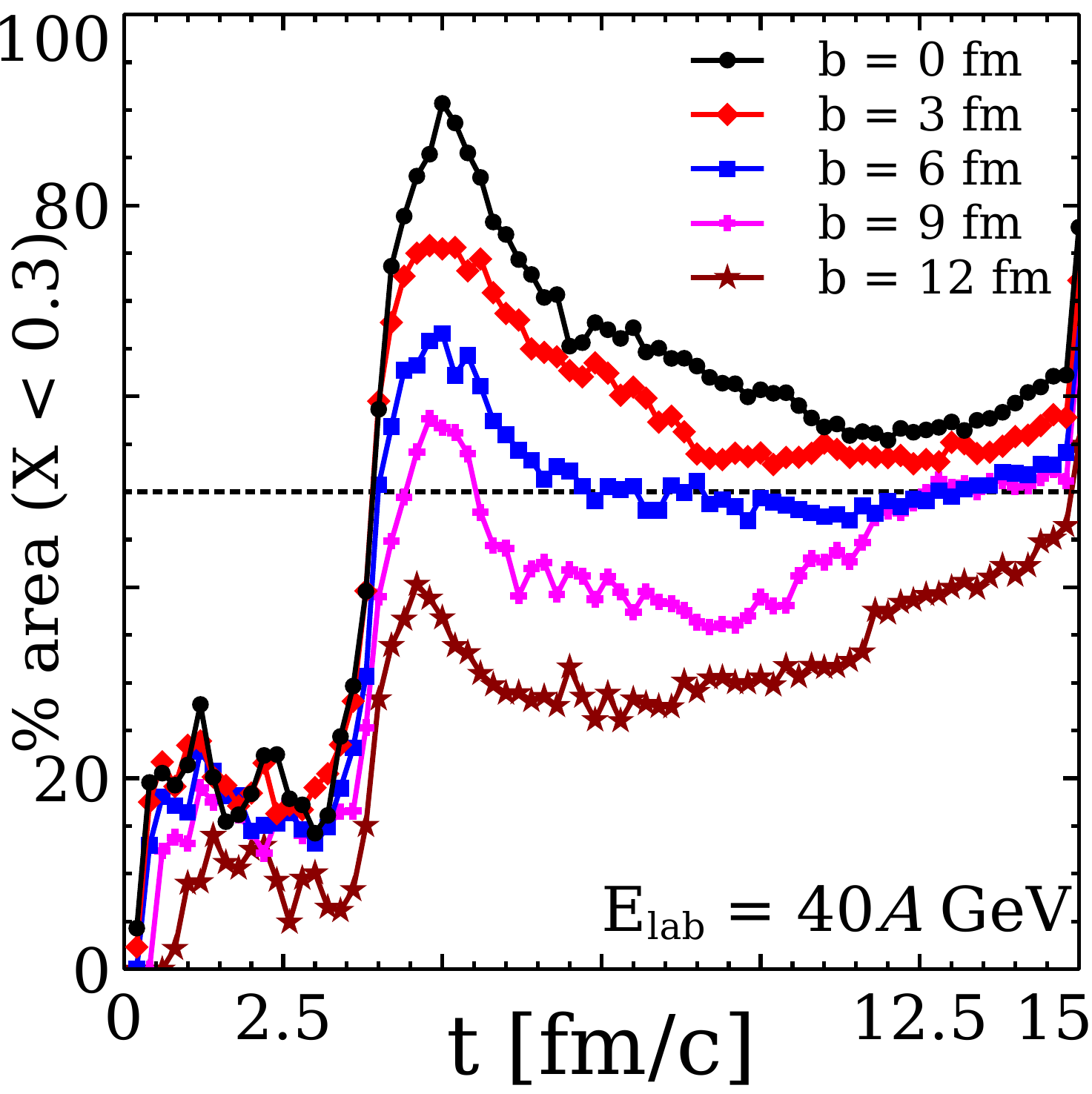} \\ 
  \includegraphics[width = 0.2\textwidth]{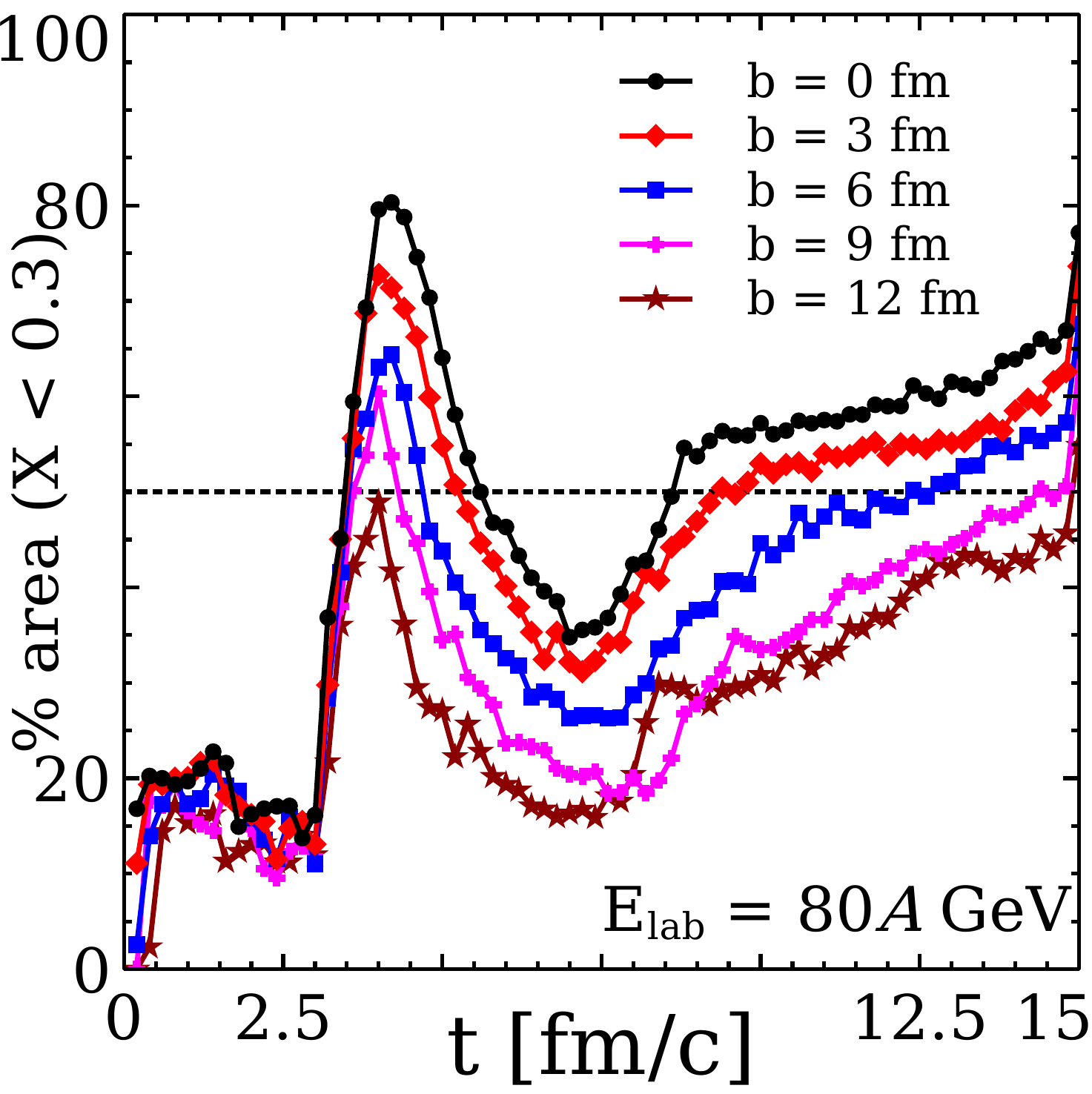}  
  \includegraphics[width = 0.2\textwidth]{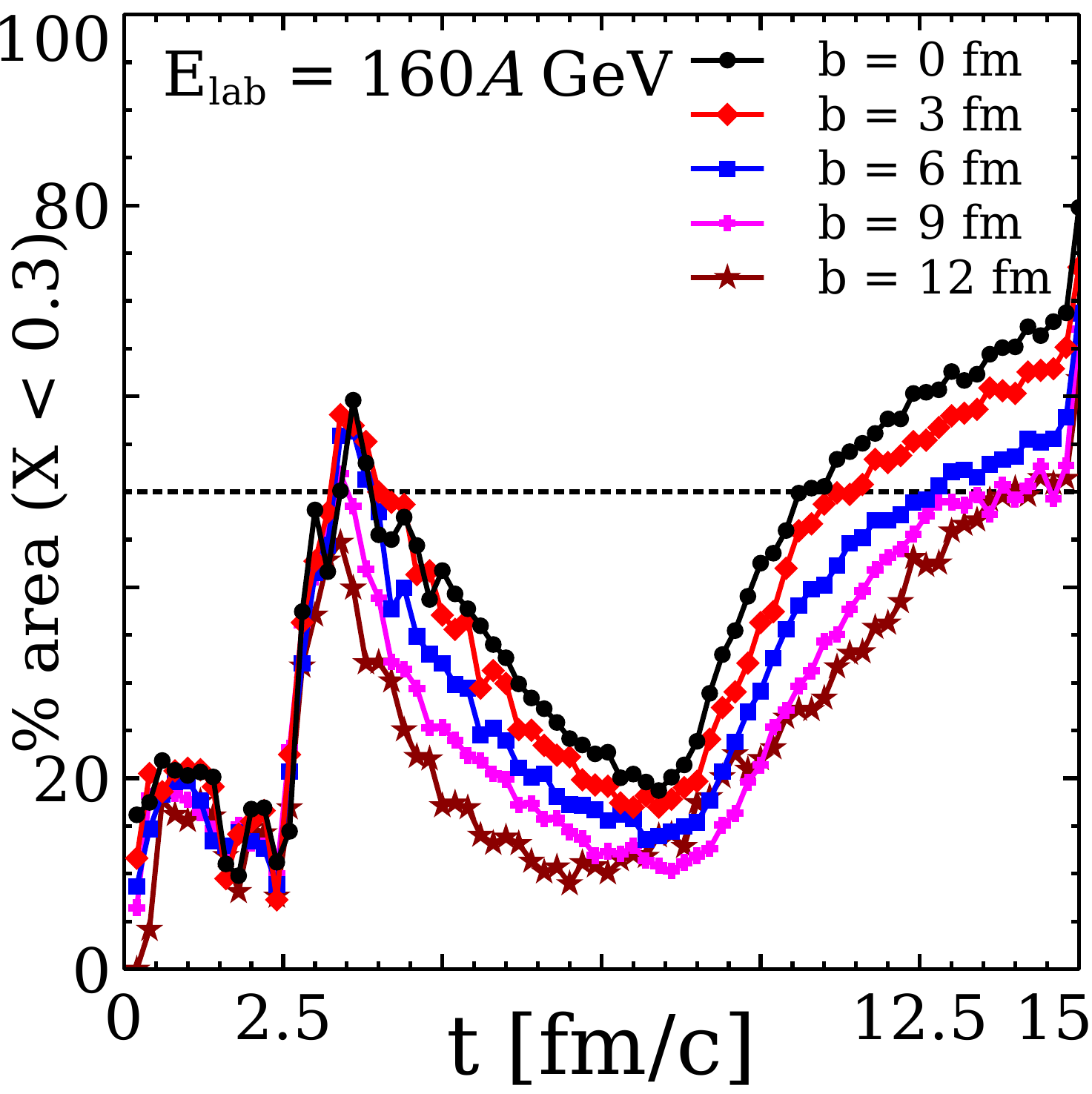}  
  \caption{Percentage of area in the event plane, where pressure anisotropy $X < 0.3$, for Au+Au collision energies $E_{lab} = 5-160\emph{A}$ GeV.}
  \label{FIG:x_area_percentages}  
\end{figure}

\begin{figure}
  \includegraphics[width = 0.4\textwidth]{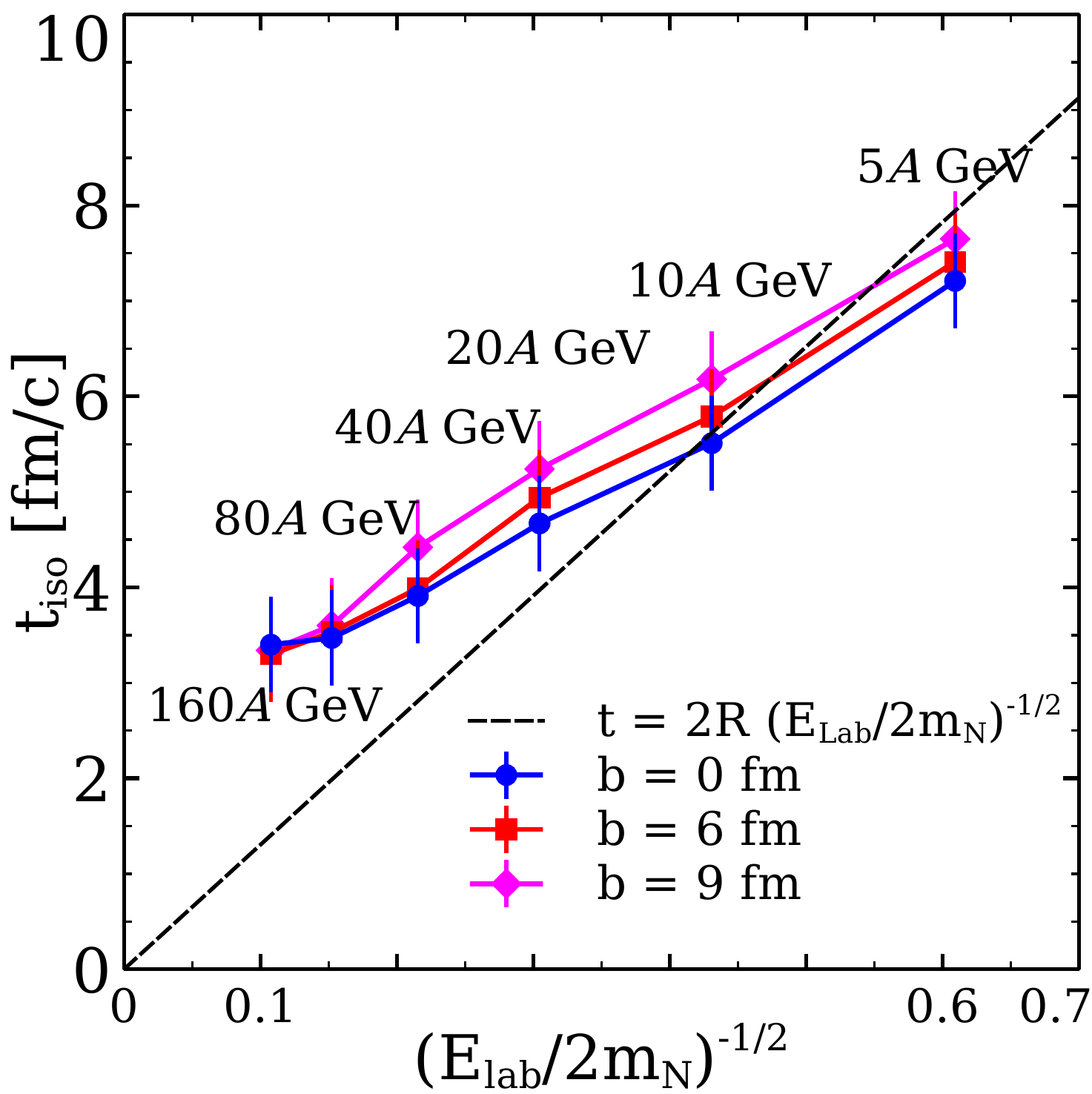}
  \caption{Isotropization time $t_{iso}$ (see definition in the text) versus energy and centrality.}
  \label{FIG:t_x_energy}
\end{figure}

\begin{figure}
  \includegraphics[width = 0.4\textwidth]{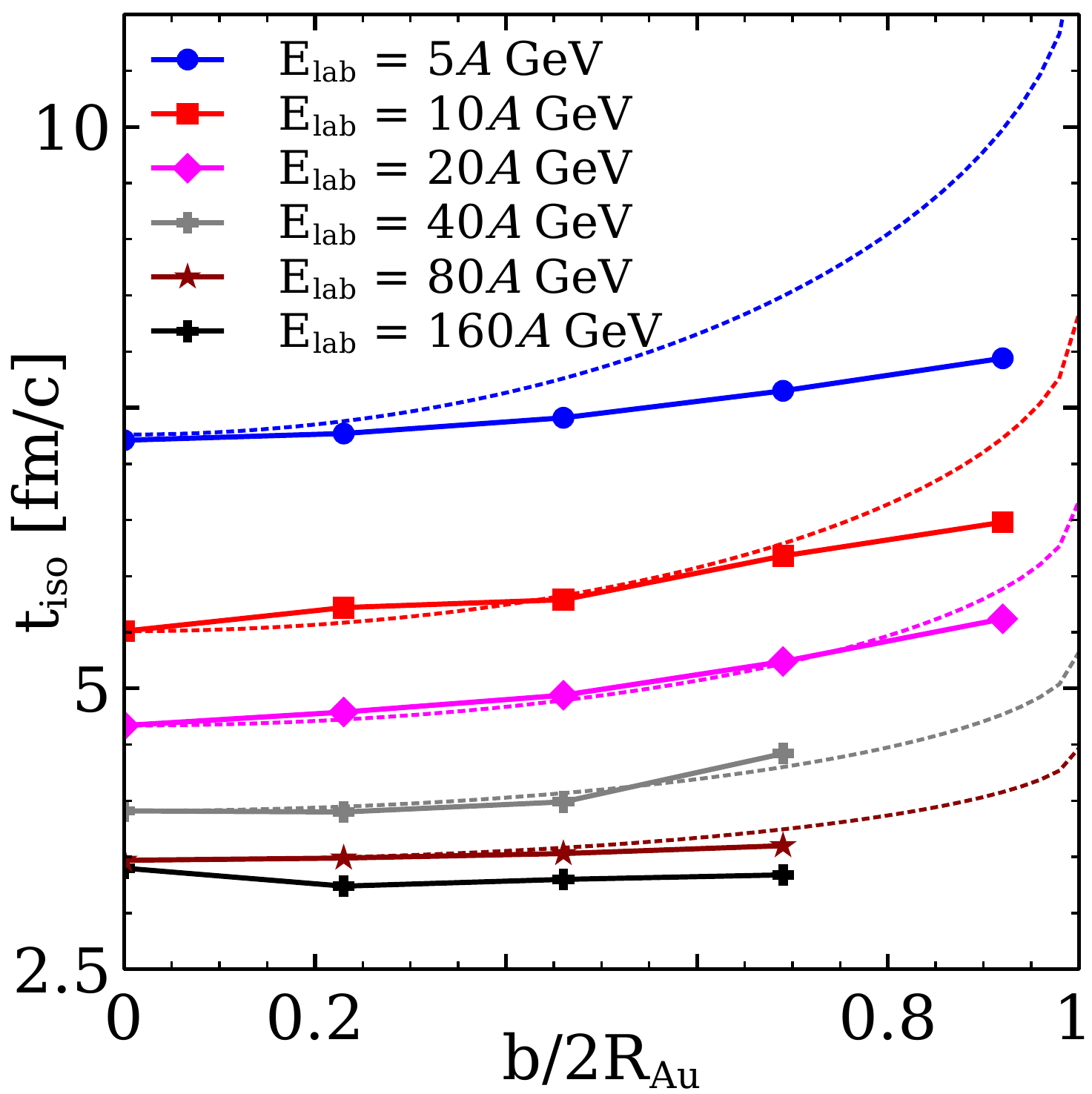}
  \caption{Isotropization time $t_{iso}$ (see definition in the text) versus centrality. Dotted lines are naive expectation from the collision geometry: $t_{iso}(b) = t_{iso}(b = 0) + \frac{R}{\gamma v} (1 - \sqrt{1 - (b/2R)^2})$.}
  \label{FIG:t_x_b}  
\end{figure}

\begin{figure*}
  \includegraphics[height = 1cm]{plots/E80b6_x_paraview_color_legend.png} \\
  \includegraphics[width = 3.5cm]{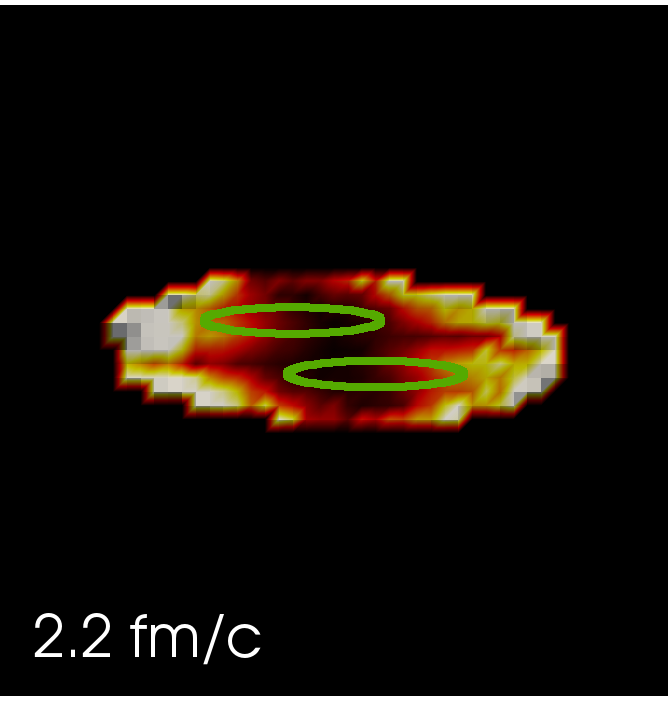}
  \includegraphics[width = 3.5cm]{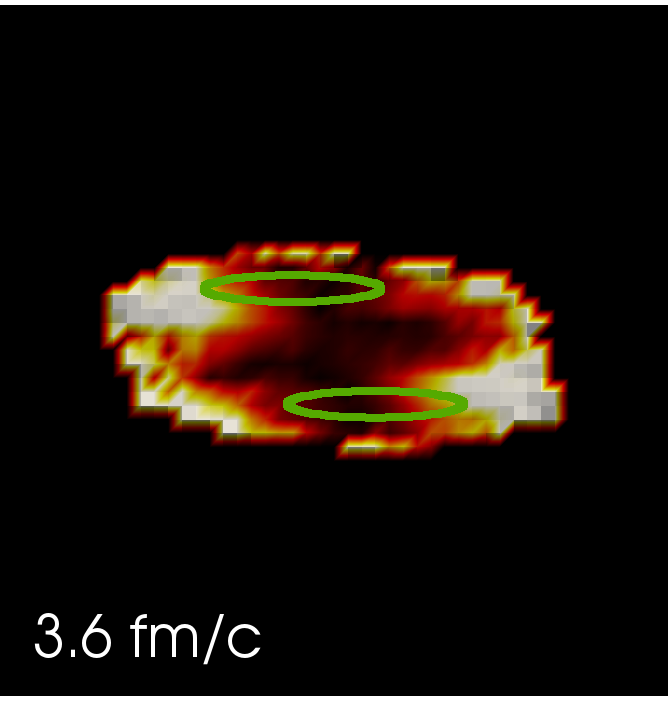}
  \includegraphics[width = 3.5cm]{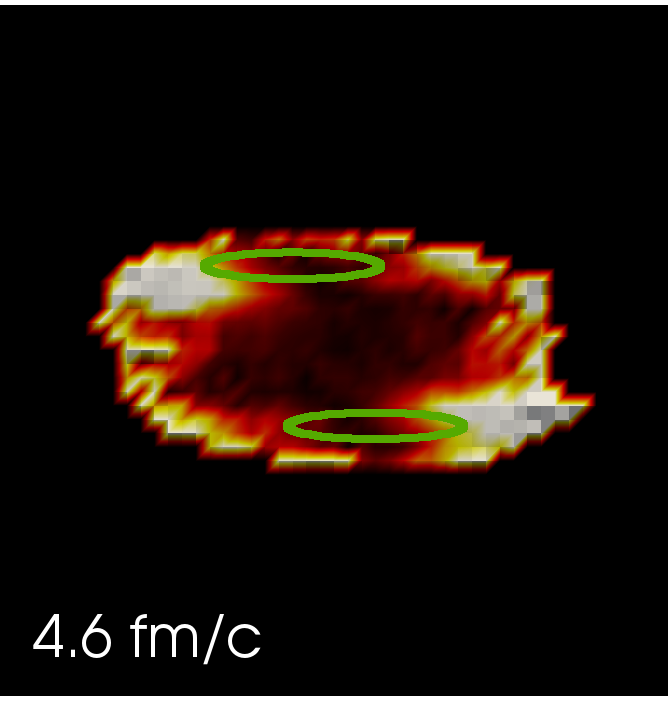}
  \includegraphics[width = 3.5cm]{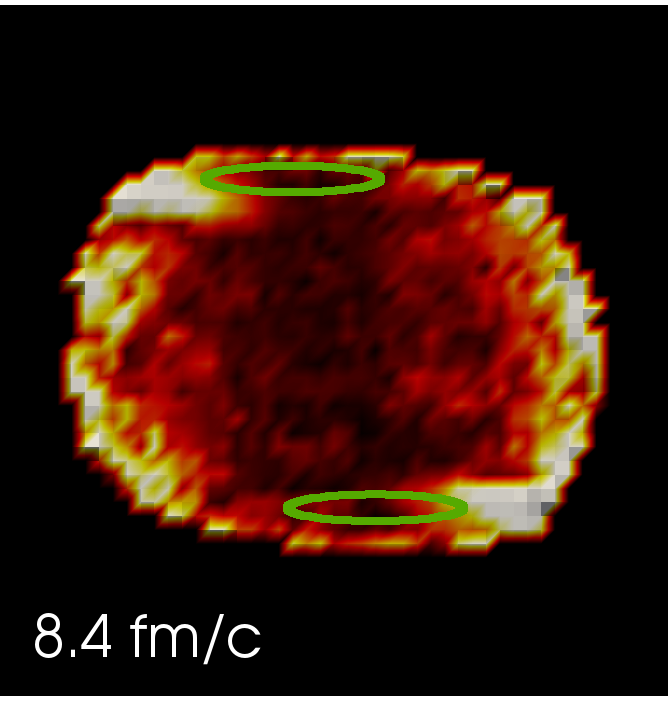}
  \includegraphics[width = 3.5cm]{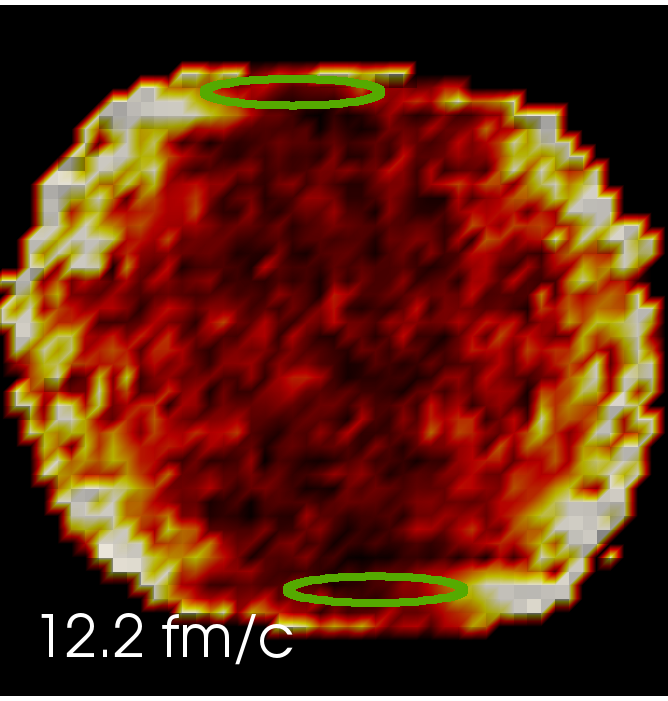}
  \caption{Space-time evolution of off-diagonality $Y = \frac{3(|T^{12}_L| + |T^{23}_L| + |T^{13}_L|)}{T^{11}_L + T^{22}_L + T^{33}_L}$ (see color scale above the Fig.) for collision energy in lab frame $E = 80\emph{A}$ GeV, centrality $b = 6$ fm. If the value of $Y$ exceeds color map maximum, it is marked with the same color as maximum. Solid lines mark the positions of the nuclei, if they wouldn't interact.}
  \label{FIG:y_paraview_space_time_evolution}
\end{figure*}

\begin{figure*}
  \includegraphics[height = 1cm]{plots/E80b6_x_paraview_color_legend.png} \\
  \includegraphics[width = 3.5cm]{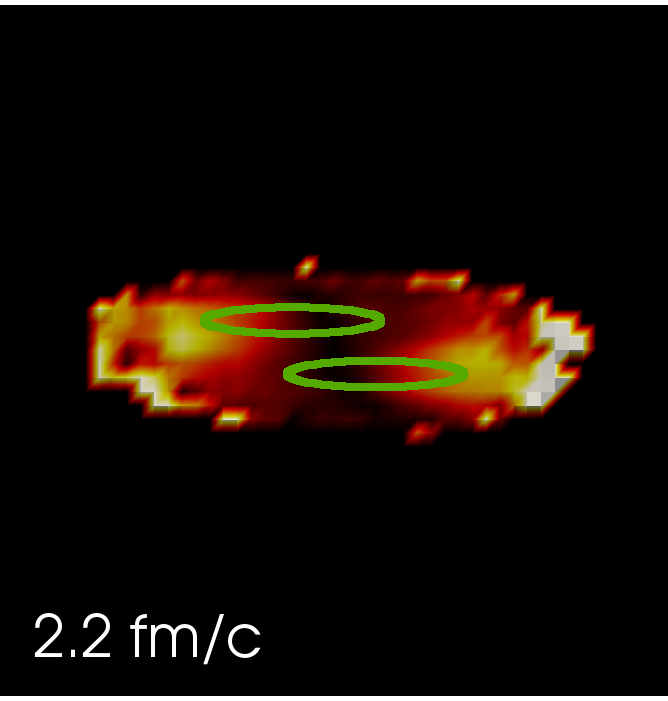}
  \includegraphics[width = 3.5cm]{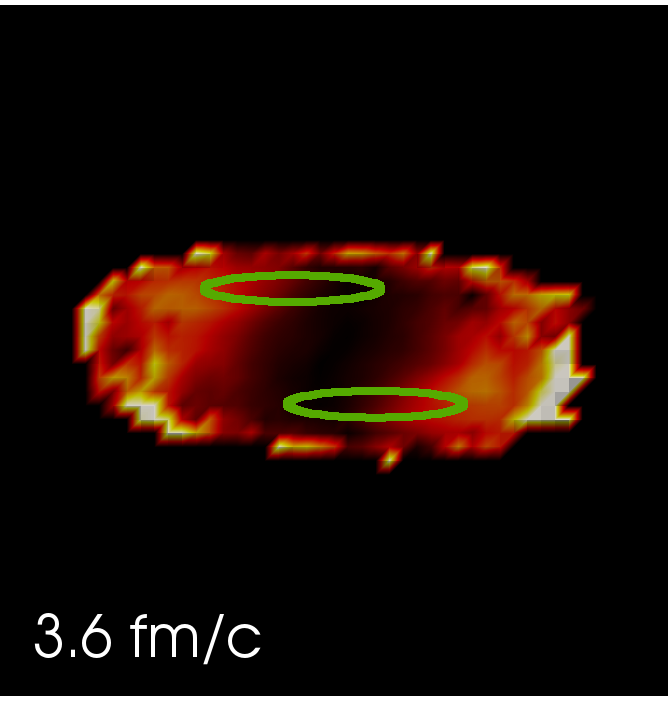}
  \includegraphics[width = 3.5cm]{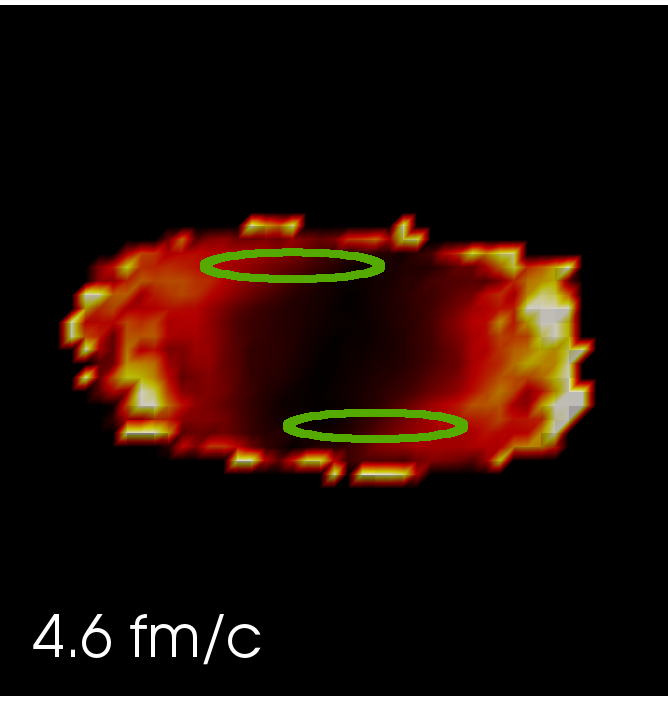}
  \includegraphics[width = 3.5cm]{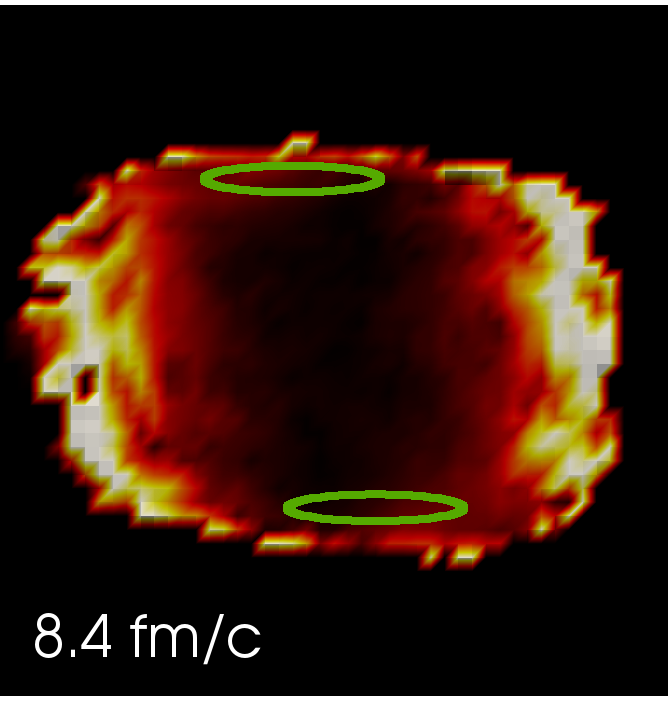}
  \includegraphics[width = 3.5cm]{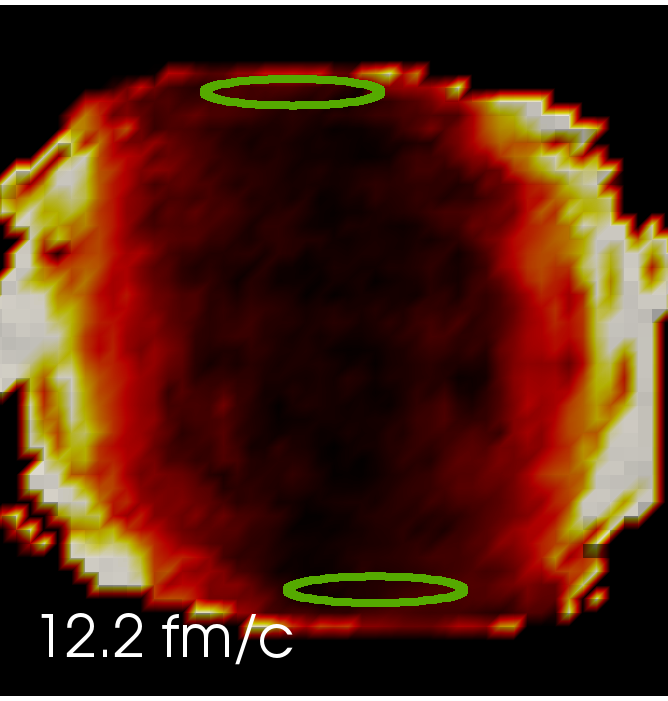}
  \caption{Space-time evolution of relative velocity between Landau and Eckart frames $v_{LE} = \sqrt{(j_L^1)^2 + (j_L^2)^2 + (j_L^3)^2}/j_L^0$ (see color scale above the Fig.) for collision energy in lab frame $E = 80\emph{A}$ GeV, centrality $b = 6$ fm. If the value of $v_{LE}$ exceeds color map maximum, it is marked with the same color as maximum. Solid lines mark the positions of the nuclei, if they wouldn't interact.}
  \label{FIG:vLE_paraview_space_time_evolution}
\end{figure*}

{\bf The pressure anisotropy} $X$ satisfies the following properties: $X \le 2$ for any tensor and $X = 0$ if and only if $T^{11}_L = T^{22}_L = T^{33}_L$ as it is for ideal fluid $\Tmn$. For viscous hydrodynamics it is necessary that $X \ll 1$. We consider $X_{crit} = 0.3$ as a limiting value, when viscous hydrodynamics is still applicable. Changing $X_{crit}$ to 0.4 leaves us with qualitatively the same results and conclusions. Fig. \ref{FIG:x_paraview_space_time_evolution} gives a qualitative impression of the space-time evolution of the pressure anisotropy. Even though the figure shows a particular energy of $E_{lab} = 80\emph{A}$ GeV and centrality $b = 6$ fm, some features are universal for all energies and centralities that we considered:
\begin{itemize}[noitemsep]
\item On the borders of the expanding system the anisotropy is always high, thus these regions are never consistent with viscous hydrodynamics.
\item After some moment of time a relatively isotropized central region rapidly expands and never disappears completely during the time evolution
\end{itemize}

To make quantitative statements let us consider the evolution of the pressure anisotropy at several points along the x axis (z = 0) versus time. This is shown in Fig.~\ref{FIG:x_vs_t_few_space_points}. One can see that in the beginning the anisotropy is almost maximal, then it rapidly decreases and never rises too much again. As we have shown in the previous section, $N_{ev} = 1000$ is enough to suppress the anisotropy due to statistics, so the behaviour of $X$ in Fig.~\ref{FIG:x_vs_t_few_space_points} is dominated by physics. However, the impact of statistical fluctuations can be observed already at $t > 5$ fm/c: $X$ starts to fluctuate in space and time. One can see this both in Fig.~\ref{FIG:x_paraview_space_time_evolution} at later times and in Fig.~\ref{FIG:x_vs_t_few_space_points}.

After looking at the space-time evolution of the anisotropy in 2D as in Fig.~\ref{FIG:x_paraview_space_time_evolution} for different energies, one gets the impression that there is a special moment $t_{iso}$ for each energy and centrality, before which pressures are highly anisotropic in the whole event plane and after which there emerges a considerable isotropic region. To quantify this feeling we consider the ratio of area, where $X < 0.3$ to the total area versus time. From Fig.~\ref{FIG:x_area_percentages} one can see that there is indeed a steep rize of isotropized area at some moment for every considered energy and centrality. Let us define the \emph{isotropization time} $t_{iso}$ such that more than 50\% of the area has $X < 0.3$ at $t = t_{iso}$. The behaviour of this isotropization time versus energy and centrality is compared to the geometrical criterion in Fig.~\ref{FIG:t_x_energy}. One can see that the isotropization time decreases with energy, but remains larger than the geometrical criterion for all energies except $5\emph{A}$~GeV. It is interesting to note that the isotropization time differs with centrality: for larger impact parameters $b$ it slightly increases. One might assume that this has a pure geometrical reason: $t_0 = 0$ is chosen in UrQMD as a moment when nuclei touch each other in a central collision. However, for peripheral collisions the nuclei will only touch at $t_0(b) = \frac{R}{\gamma v} (1 - \sqrt{1 - (b/2R)^2})$. In Fig.~\ref{FIG:t_x_b} one can see that this naive expectation yields the right trend: $t_{iso}$ rises with centrality and the rise is smaller for higher energies. However, quantitatively it overestimates $t_{iso}$ for large impact parameters.

Let us compare our findings to the study by Bravina et al. \cite{Bravina:2008ra}, where one central cell of $(5 \times 5 \times 5)$ fm$^3$ was chosen to study the pressure anisotropy of the energy-momentum tensor in Au+Au collisions. An isotropization time was  defined, and it did not change significantly after zooming the central cell to $(1 \times 1 \times 1)$ fm$^3$. This allowed for the conclusion that isotropization happens rapidly in a large volume. The isotropization time determined in this study also decreases with collision energy. All of these results are confirmed in the present study. However, our isotropization times are smaller than the ones obtained by Bravina et al. There are two possible reasons for that. First, we do not count spectators for constructing pressure anisotropies. Second, the criterion for isotropization we use is less strict: while we only require at least 50\% of event plane area to have $X < 0.3$, the central cell study demands $p_z/p_x - 1 < 0.1$ in the whole cell, which corresponds to $X < 0.065$.

{\bf The off-diagonality} $Y = \frac{3(|T^{12}_L| + |T^{23}_L| + |T^{13}_L|)}{T^{11}_L + T^{22}_L + T^{33}_L}$ characterizes the size of the off-diagonal components of the stress tensor compared to the pressure. In ideal hydrodynamics $Y = 0$. For the applicability of viscous hydrodynamics it is necessary that $Y \ll 1$. We consider $Y_{crit} = 0.3$ as a value, after which viscous hydrodynamics is hardly applicable. An example for the space-time evolution of $Y$ is given in Fig.~\ref{FIG:y_paraview_space_time_evolution}. From this Figure it can be seen that in the central region $Y$ is always small. On the boundaries $Y$ is typically large due to statistical effects. A quantitative study similar to the study of the pressure isotropy $X$ shows that for all considered energies and centralities more than 80\% of the event plane area have $Y < 0.3$ for the whole time of the evolution. 

{\bf The relative velocity between Landau and Eckart frames} for baryon charge $v_{LE}$ is shown in Fig.~\ref{FIG:vLE_paraview_space_time_evolution}. At high enough statistics the relative velocity between Eckart and Landau frames is not an important factor. It is significant only on the borders of the system, where the density is small and statistical effects play a role. But in all the rest of the volume, for all the considered time evolution it remains small.

{\bf The effect of momentum-space cuts}
\begin{figure}
  \includegraphics[height = 1cm]{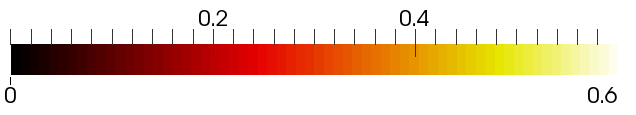} \\
  \includegraphics[width = 8cm]{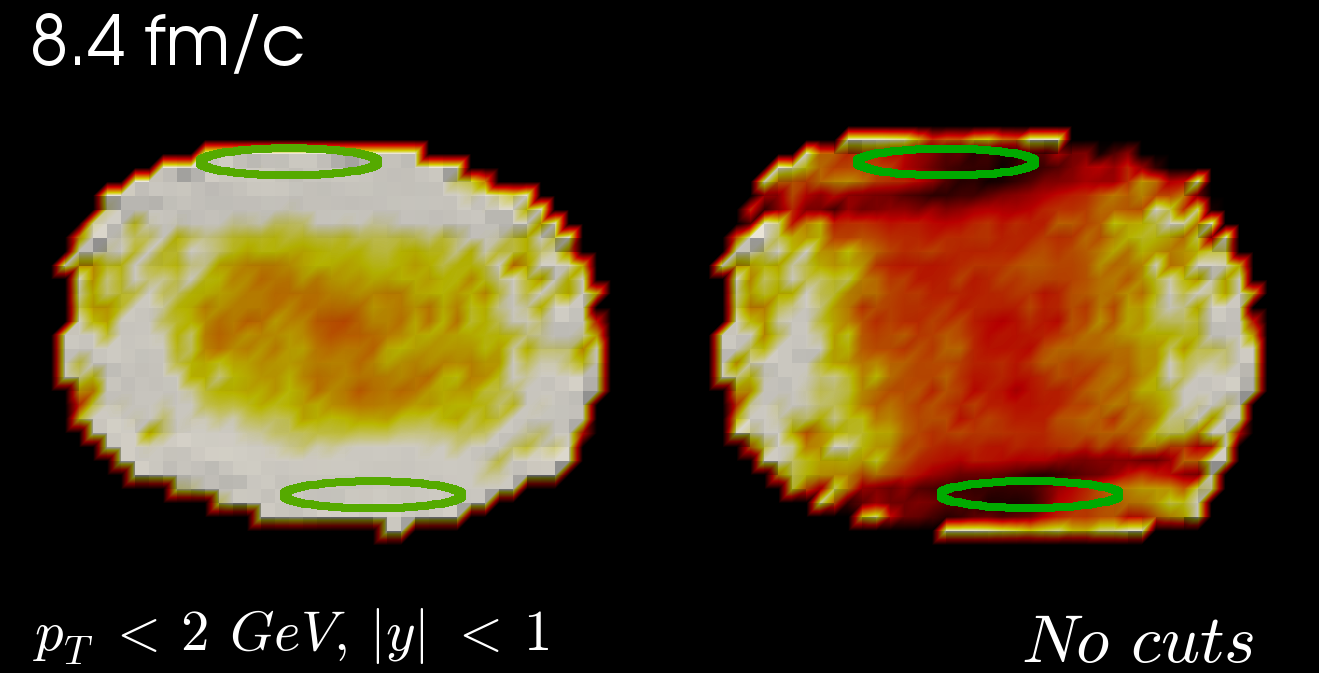}
  \includegraphics[width = 8cm]{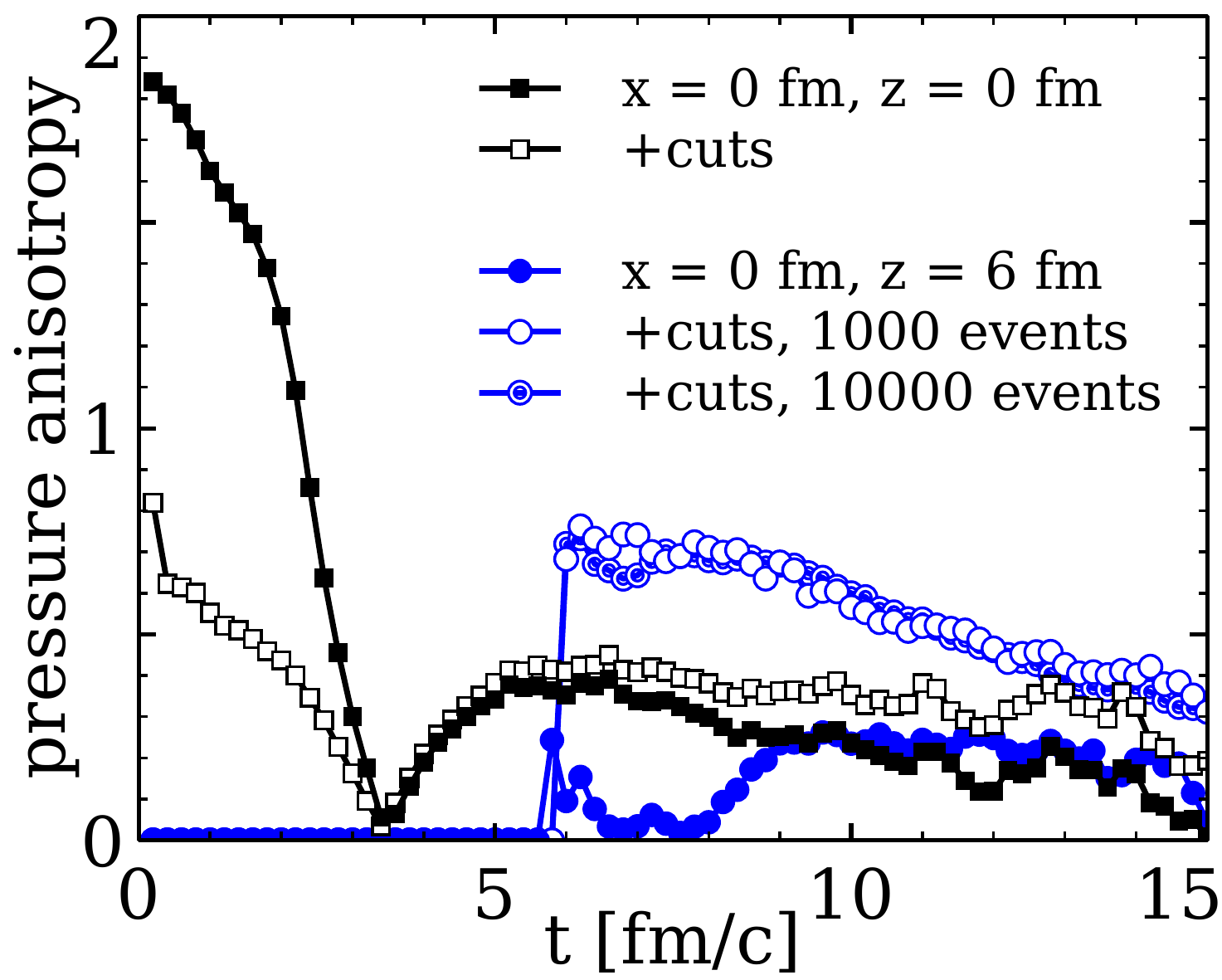}
   \caption{The effect of $p_T < 2$ GeV and $|y| < 1$ cuts on the space distribution of pressure anisotropy $X = \frac{|T_L^{11}-T_L^{22}|+|T_L^{22}-T_L^{33}|+|T_L^{33}-T_L^{11}|}{T_L^{11}+T_L^{22}+T_L^{33}}$ (see color scale above the Fig.) for collision energy in lab frame $E = 80\emph{A}$ GeV, centrality $b = 6$ fm. If the value of $X$ exceeds color map maximum, it is marked with the same color as maximum. Solid lines mark the positions of the nuclei, if they wouldn't interact.}
  \label{FIG:X_paraview_cuts_effect}
\end{figure} 

Previously, we included all participants into $\Tmn$ calculation. However, it is generally believed that soft particles at midrapidity thermalize faster, therefore it might be insightful to impose cuts in momentum space, if one wants to obtain a more isotropic $\Tmn$. On the other hand, applying tranverse momentum $p_T$ and rapidity $y$ cuts on a perfectly symmetric distribution results in an asymmetry. In addition, these cuts decrease the statistics, which leads to an increase of the anisotropy. To study the effect of the kinematic cuts on the space distribution of the pressure anisotropy $X$, we constructed $\Tmn$ only from particles with $p_T < 2$ GeV and $|y| < 1$. The effect of cuts on $X$ over space is shown in Fig. \ref{FIG:X_paraview_cuts_effect}. We made sure that the statistical effect does not play a significant role by checking that results do not change after increasing the number of events from $N_{ev} = 10^3$ to $10^4$. The large anisotropy at very early times decreases after imposing cuts. But the anisotropy at later times is strikingly larger with cuts, first of all in the regions behind the nuclei. While without cuts we were able to introduce an ''isotropization time'', when $X$ is smaller than $0.3$ at more than 50\% of the area, with the cuts the anisotropy is so high all over the space that the isotropization time cannot be introduced anymore.

\section{Conclusions}

We have studied the energy-momentum tensor from the UrQMD transport approach locally in space and time and quantified its deviation from the ideal fluid form with two numbers: the pressure anisotropy $X$ and the off-diagonality $Y$. First of all, we have shown that $X$ and $Y$ depend on the number of UrQMD events $N_{ev}$ used to construct $\Tmn$. Low statistics implies large deviations, even if the underlying distribution function is completely thermal and isotropic. An initial state constructed from less than few hundred events (or few hundred testparticles equivalently) is bound to deviate strongly from the ideal fluid form. The off-diagonality appears to be mostly produced by this statistics effect. For large statistics $Y$ tends to be small in all the collision region at all times. The pressure anisotropy does not vanish at large statistics, it is a physical effect related to the anisotropy of the underlying distribution function $\mathit{f}(\vec{r},\vec{p})$. As a consequence, the initial state from UrQMD with enough statistics is suitable for anisotropic hydrodynamics.

Unfortunately, all the results depend on the smearing parameter $\sigma$. With larger $\sigma$ isotropization is reached later, but the degree of isotropization is higher. From a practical point of view that means that selecting large $\sigma$ one has to take larger fluidization time. This is in agreement with conclusion of \cite{Petersen:2010zt} that larger fluidization time should be taken for larger $\sigma$ to obtain the same pion yield. However, strictly speaking, the physical limit is $\sigma \to 0, \, \sigma^3 \rho N_{ev} \to \infty$. We found that at small $\sigma$ the isotropization time approaches to the time of geometrical criterion $t_{geom} = 2R (E_{lab}/2m_N)^{-1/2}$, so the geometrical criterion is partly justified, but only in the above mentioned limit. In the existing models the smearing parameters and statistics are such that at $t_{geom}$ the anisotropies are very high.

For the pressure anisotropy $X$ it was observed, that it exhibits a similar pattern for all the considered collision energies and centralities: there is a narrow interval of time, when it rapidly drops in a considerable volume. This feature allowed us to introduce and study the isotropization time $t_{iso}$. It decreases with collision energy, following the same trend as the geometrical overlap time. Based on this finding a new fluidization criterion is suggested: $t_{iso} = t_{geom}(E) + \Delta t_0 (\sigma)$, where $\Delta t_0$ depends only on the Gaussian smearing $\sigma$ and can be determined from Fig.~\ref{FIG:t_iso_sensitivity_sigma}. We observed a slight dependence of isotropization time on centrality: it increases with impact parameter, but the slope of this increase becomes smaller and smaller for higher energies. This behaviour has a simple geometrical interpretation.

Testing the smallness of the bulk pressure compared to pressure left beyond the scope of this article. As we have mentioned, the extracted bulk pressure depends on the equation of state, therefore this check has to be performed with a set of EoS used in the literature. Since in this article we restricted ourselves to testing the weak consistency, the next possible step is to test locally the fulfilment of the strong consistency.

\begin{acknowledgments}
  This work was supported by the Helmholtz International Center for
  the Facility for Antiproton and Ion Research (HIC for FAIR) within
  the framework of the Landes-Offensive zur Entwicklung
  Wissenschaftlich-Oekonomischer Exzellenz (LOEWE) program launched by
  the State of Hesse. HP acknowledges funding of a Helmholtz
  Young Investigator Group VH-NG-822 from the Helmholtz Association
  and GSI. DO acknowledges support of the Deutsche Telekom Stiftung.
  Computational resources have been provided by the Center for Scientific
  Computing (CSC) at the Goethe University of Frankfurt. We thank Marlene Nahrgang
  and Harri Niemi for reading the manuscript and useful comments.
\end{acknowledgments}

\section{Appendix: $\Tmn$ and $\jmu$ from particles}

The general formula for calculating $\Tmn$ and $\jmu$ from particles is
\begin{align}
\Tmn_{init}(r) &=& \int \frac{p^{\mu} p^{\nu}}{p^0} \mathit{f}(\vec{r}, \vec{p}) d^3p \\
\jmu_{init}(r) &=& \int \frac{p^{\mu}}{p^0} \mathit{f}(\vec{r}, \vec{p}) d^3p \nonumber
\end{align}
Here $\mathit{f}$ is the 1-particle distribution function. For a discrete set of particles it is
\begin{align}
\mathit{f}(\vec{r}, \vec{p}) = \sum_{part}  \delta^3(\vec{p} - \vec{p_{part}}) \delta^3 (\vec{r} - \vec{r_{part}})
\end{align}
For numerical calculations the spatial dependence of the distribution function needs to be smeared in some way, therefore the delta-function is substituted by some smearing kernel $K(\vec{r} - \vec{r_{part}})$. The smearing kernel should satisfy three conditions:
$K(\vec{r})d^3r$ should be Lorentz scalar, it should be normalized as $\int K(\vec{r})d^3r = 1$, and it should approach the delta-function for smearing parameters approaching to zero. Surprisingly, all current popular choices of $K$ are such that $K(\vec{r})d^3r$ is not a Lorentz scalar:
\begin{itemize}[noitemsep]
\item Cell averaging: $K(\vec{r}) = \begin{cases} 1/\Delta V, \, \vec{r} \in \Delta V \\
                                                  0, \text{otherwise} \end{cases}$. Here $K(\vec{r})d^3r$ is not a Lorentz scalar, since volume $V$ is not contracted.
\item Gaussian averaging with Lorentz contraction in z direction:
                      $K(\vec{r}) = N exp\left(-\frac{x^2 + y^2 + \gamma_z^2 z^2}{2 \sigma^2} \right)$ is behaving properly only under boosts in z direction.
\item Gaussian in $x$, $y$, $\eta$ coordinates also behaves properly only under boosts in z direction.
\end{itemize}
Further we derive a simple, not too computationally demanding kernel, which satisfies all aforementioned conditions, in particular $K(\vec{r})d^3r$ being a Lorentz scalar. In the rest frame of the particle we take a Gaussian: $K(\vec{r_{rest}}) = (2\pi \sigma^2)^{-3/2} exp(-\vec{r^2_{rest}}/2\sigma^2)$. If $\Delta x'^{\mu}$ are the coordinates in the computational frame and $\Delta x^{\mu}$ are the coordinates in the rest frame then $\Delta x^{\mu} = \Lambda^{\mu}_{\nu} \Delta x'^{\nu}$, where $\Lambda$ is the matrix of the Lorentz transformation. We want to consider a Gaussian at fixed time in the computational frame, so $\Delta x'^0 = 0$. Denoting the spatial part of the particle 4-velocity as $\vec{u} = \gamma \vec{\beta}$, where $\vec{\beta}$ is 3-velocity and $\gamma = (1 - \vec{\beta}^2)^{-1/2}$, one obtains
\begin{align}
\Delta x^i = \Lambda^{i}_{j} \Delta x'^{j} \\
\Lambda^i_j = \delta^i_j + (u^i u_j)/(1 + \gamma) \\
(\Delta x^i)^2 = \Lambda^{i}_{j} \Delta x'^{j} \Lambda^{i}_{k} \Delta x'^{k} \\
\Lambda^{i}_{j} \Lambda^{i}_{k} = \delta_{jk} + u_j u_k \\
(\Delta \vec{x})^2 = (\Delta \vec{x'})^2 + (\Delta \vec{x'} \cdot \vec{u})^2 \\
K(\vec{r}) = N_0 (2\pi \sigma^2)^{-3/2} exp\left(-\frac{\vec{r}^2 + (\vec{r} \cdot \vec{u})^2}{2 \sigma^2} \right)
\end{align}
To determine the normalization factor $N_0$, one can use that $\int \left( \prod_{i=1}^{n} dx_i \right) e^{x_i A^{ij} x_j}  = \pi^{n/2} \, (det A)^{-1/2}$. The determinant $det(\Lambda^{i}_{j} \Lambda^{i}_{k}) = det(\delta_{jk} + u_j u_k) = \gamma^2$, so $\int K(\vec{r})d^3r = N_0/\gamma = 1$ and therefore $N_0 = \gamma$. Finally we obtain
\begin{align}
K(\vec{r}) = (2\pi \sigma^2)^{-3/2} \gamma \, exp\left(-\frac{\vec{r}^2 + (\vec{r} \cdot \vec{u})^2}{2 \sigma^2} \right)
\end{align}
One can check that this formula turns into the commonly used one if $\vec{u} = \gamma_z(0, 0, \beta_z)$. Since $1 + (\gamma_z \beta_z)^2 = \gamma_z^2$ in this case, we immediately retain the usual smearing kernel. Scalar products do not change under rotations, so the expression in the exponential does not change under rotations either. Rotating to the frame, where z axis is parallel to $\vec{u}$ one can see that our kernel is nothing else, but a proper Lorentz contraction in the direction of motion.

\end{document}